\documentclass[fleqn,usenatbib]{mnras}

\usepackage[T1]{fontenc}
\usepackage{booktabs}
\DeclareRobustCommand{\VAN}[3]{#2}
\let\VANthebibliography\thebibliography
\def\thebibliography{\DeclareRobustCommand{\VAN}[3]{##3}\VANthebibliography}

\usepackage{lmodern}
\usepackage{graphicx}	
\usepackage{amsmath}	
\usepackage{amssymb}	
\usepackage[table]{xcolor}
\usepackage{multirow}
\usepackage{newtxtext, newtxmath}
\usepackage{mathtools}
\usepackage{booktabs,caption}
\usepackage[flushleft]{threeparttable}

\definecolor{ttblue}{RGB}{91,194,224}
\definecolor{colorbanda}{HTML}{003f5c}
\definecolor{colorbandb}{HTML}{7a5195}
\definecolor{colorbandc}{HTML}{ffa600}
\definecolor{tableShade}{gray}{0.9}

\defcitealias{Altamura2023}{Paper~I}

\title[Entropy plateaus at a characteristic halo mass]{Entropy plateaus can emerge from gas replacement at a characteristic halo mass in simulated groups and clusters of galaxies}

\author[E. Altamura et al.]{
Edoardo Altamura,$^{1}$
Scott T. Kay,$^{1}$\thanks{E-mail: \href{mailto:scott.kay@manchester.ac.uk}{scott.kay@manchester.ac.uk}}
Joop Schaye,$^{2}$ 
Ian G. McCarthy,$^{3}$ 
and
Matthieu Schaller$^{2, 4}$ 
\\
$^{1}$Jodrell Bank Centre for Astrophysics, Department of Physics and Astronomy, The University of Manchester, Oxford Road, Manchester M13 9PL, UK\\
$^{2}$Leiden Observatory, Leiden University, PO Box 9513, NL-2300 RA Leiden, the Netherlands\\
$^{3}$Astrophysics Research Institute, Liverpool John Moores University, Liverpool L3 5RF, UK\\
$^{4}$Lorentz Institute for Theoretical Physics, Leiden University, PO box 9506, 2300 RA Leiden, the Netherlands\\
}

\date{Accepted XXX. Received YYY; in original form ZZZ}

\pubyear{2025}

\begin{document}
\label{firstpage}
\pagerange{\pageref{firstpage}--\pageref{lastpage}}
\maketitle

\begin{abstract}
The evolution of the intergalactic medium (IGM) is influenced by gravitational collapse, radiative cooling, and baryonic feedback. Using cosmological hydrodynamic zoom-in simulations of a $8.83 \times 10^{12}$ M$_\odot$ group and a $2.92 \times 10^{14}$ M$_\odot$ cluster at $z=0$, we investigate the emergence of entropy plateaus and their connection to feedback mechanisms. This set-up uses the SWIFT-EAGLE model with three resolutions, down to an initial particle gas mass of $2.29 \times 10^5$ M$_\odot$ and $1.23 \times 10^6$ M$_\odot$ for dark matter. We find that, when halos reach the characteristic mass of $\sim 10^{12}$ M$_{\odot}$, their entropy profiles flatten at the virial radius, marking a transition from supernova to AGN feedback-driven regulation. As halos grow into groups ($\sim 10^{13}$ M$_{\odot}$), the entropy plateau extends inward and isentropic cores form in massive systems ($\sim 10^{14}$ M$_{\odot}$). By tracking the Lagrangian history of gas particles, we demonstrate that this entropy buildup is primarily driven by AGN feedback, which efficiently removes low-entropy gas from progenitors of groups and clusters, redistributing it throughout the IGM before falling into the core. Recent observations of X-GAP groups reveal large entropy excesses and plateaus, in line with our findings and in contrast to the power-law-like profiles of most previous observations. While entropy plateaus and large entropy excesses may be observationally confirmed in unbiased samples, reproducing the full diversity of entropy profiles remains an outstanding challenge for next-generation feedback models. Our results suggest that current feedback models may be overly efficient in expelling low-entropy gas from the potential cool-core progenitors, disrupting the balance between heating and cooling required for long-lived cool cores.
\end{abstract}




\section{Introduction}

The evolution of galaxies through cosmic time is deeply intertwined with the complex life cycle of baryonic matter, composed of gas, dust, stars and black holes. The baryon cycle describes the gas transfer between the galaxies and their surroundings (Inter-galactic Medium, IGM). It drives the growth of galactic structures and regulates the thermodynamic properties of the IGM. The pathways of the baryon cycle are characterised by inter-dependent processes, such as star formation, outflows from active galactic nuclei (AGN), supernova (SN) explosions, radiative cooling, thermal instability, and the gravity-driven accretion of gas back onto galaxies \citep[see][for a review]{DONAHUE20221}.

The transport of gas parcels in an otherwise hydrostatic IGM is governed by buoyancy, whereby cold and dense gas moves towards the centre of the gravitational potential. In contrast, hot and diffuse gas moves away from the centre and stabilises near gas with similar entropy. In hydrodynamics, this process can be probed with the value of the pseudo-entropy, defined as $K=k_{\rm B}T / n_e^{(\gamma - 1)}$ in terms of the temperature $T$, the electron number density $n_e$, and the adiabatic index $\gamma$ of the ionised gas, with $k_{\rm B}$ the Boltzmann constant\footnote{The pseudo-entropy $K$, referred to as \textit{entropy} in this work, is related to the classical thermodynamic entropy as $S\propto\log K$. We direct the reader to \cite{entropy_intro_bower_1997} for the derivation of $K$ for a mono-atomic gas, which has $\gamma=5/3$ and hence $K=k_{\rm B}T/n_e^{2/3}$.}. Low-entropy gas sinks to lower `altitude', i.e. towards the centre, while high-entropy gas buoyantly rises towards higher altitudes, generating convective motions that tend to align the entropy gradient with the gradient of the gravitational potential \citep[Schwarzschild criterion, see][]{1958ses..book.....S, 1965ApJ...142..229L, 1966ApJ...144..201T}.

These conditions can be recreated in \textit{non-radiative} simulations of groups and clusters and lead to self-similar power-law-like entropy profiles, characterised by low entropy levels in the core and higher entropy at the virial radius \citep{2001ApJ...546...63T, vkb_2005}. X-ray observations of groups and clusters with masses $10^{13}-10^{15}$ M$_\odot$ similarly reveal entropy profiles with a median power-law behaviour outside the core \citep{entropy_profiles_sun2009, entropy_profiles_pratt2010}.

Here, the radial entropy distribution appears to follow a universal power-law profile \citep{2012MNRAS.427L..45W} with a logarithmic slope of $d\log K / d\log r \equiv \alpha_K\approx 1.21$ \citep{vkb_2005}. The fundamental physics underpinning the universality of this feature has not been fully explored. Two explanations attempt to describe the steady-state entropy profile of the IGM gas and the dark matter \citep{2007MNRAS.376.1327F, 2012ApJ...756..100H}. The first uses statistical physics to maximise the entropy functional for a self-gravitating particle ensemble in steady state to obtain power-law/polytropic-like profiles \citep{1993PhLA..174..384P, 2010MNRAS.406.2678H, 2015MNRAS.452..944B}. The second approach suggests that the power-law index in macroscopic entropy profiles arises from dynamical effects, like radial orbit instability, phase mixing, or violent relaxation that occur during the halo-assembly history \citep[e.g.][]{2003MNRAS.341..927K, 2009ApJ...690..102H, 2022MNRAS.513..573D}. 

Notwithstanding the self-similar scaling close to the virial radius, groups and clusters show a remarkably diverse entropy distribution in their cores, where baryonic processes are important \citep{2022MNRAS.513..573D}. In fact, X-ray observations of groups \citep{entropy_profiles_sun2009} and the REXCESS clusters in \cite{entropy_profiles_pratt2010} confirmed that significant variation in the entropy profiles is present near the core. Groups and clusters with a small ($\approx 3~\mathrm{keV\, cm^2}$) excess entropy in the core over a power-law baseline are classified as cool-cores (CC), while large ($\approx 75~\mathrm{keV\, cm^2}$) entropy excess characterises non-cool-core (NCC) objects.

According to X-ray observations, the CC population of clusters is measured to be up to 64 per cent of the total -- analysing an X-ray-selected sample of clusters, \cite{2017ApJ...843...76A} and \cite{2017MNRAS.468.1917R} found that the CC population was 44-64 per cent of the total, depending on the CC definition. However, these samples are affected by the cool-core bias \citep{2011A&A...526A..79E}, which favours concentrated and X-ray-bright CCs \citep{2015ApJ...802...34L, 2017MNRAS.468.1917R}. Instead, the cool-core bias does not affect Sunyaev-Zeldovich (SZ)-selected samples as they are close to being mass-limited, showing much lower CC fractions observed by \cite{2017ApJ...843...76A, 2017MNRAS.468.1917R} (28-59 per cent) and found to be weakly dependent on redshift \citep{2017ApJ...843...28M}. On the other hand, hydrodynamic simulations predict different scenarios. In some cases, CC clusters make up the vast majority of the population (e.g., CLEF from \citealt{2007MNRAS.377..317K}, and MACSIS from \citealt{barnes_macsis}), and in other cases, most of the objects have an NCC (e.g., Cluster-EAGLE sample, \citealt{barnes_ceagle}, and the \textit{extended sample} of \citealt{Altamura2023}, henceforth \citetalias{Altamura2023}). As an intermediate case, the study of \citet{2024MNRAS.533.2656B} using the FLAMINGO simulations \citep[fiducial model, L2p8\_m9,][]{2023MNRAS.526.4978S} predicts a CC fraction similar to TNG-Cluster \citep{lehle_tng_cluster} and marginally higher than IllustrisTNG \citep{2018MNRAS.481.1809B}, particularly when using a self-similarly evolving cooling-time criterion to discriminate CC from NCC clusters. Without the time-evolving factor, however, the FLAMINGO CC fraction increases at higher redshifts, in conflict with observations of an approximately constant factor-less CC fraction \citep{2017ApJ...843...28M, 2021ApJ...918...43R}. Finally, the low-redshift CC fraction in simulations like Magneticum \textit{Box2b/hr} \citep{2015MNRAS.451.4277D, 2025A&A...694A.232G} and IllustrisTNG \citep{2018MNRAS.473.4077P} appear to be in reasonable agreement with observations when selecting objects by central electron number density and the scaled concentration parameter, but not when selecting by central entropy \citep{2018MNRAS.481.1809B}. Table~\ref{tab:resolutions-comparison-other-works} contains a compilation of studies of simulated groups and clusters quoting their sub-grid model, resolution parameters and whether enough cool cores are produced to match observations.

Over the past two decades, reproducing the CC fraction in simulations proved to be challenging for two reasons: firstly, the observed fraction of CC does not emerge \textit{ab initio} from the cosmological parameters used; secondly, the impact of feedback and cooling on group and cluster cores is non-linear and depends on several strongly correlated sub-grid parameters \citep{2011ASL.....4..204B}, complicating the fine-tuning of the model. So far, no explicit attempt has been made to control the CC fraction in cosmological simulations and theoretical predictions for observations are generally calculated \textit{a posteriori} with a sub-grid model calibrated on galaxy-related quantities \citep[e.g.][]{2015MNRAS.450.1937C}.

In \citetalias{Altamura2023}, we investigated hot gas and star fractions and entropy profiles at $z=0$ in groups and clusters, simulated with a version of the SWIFT-EAGLE model similar to that presented in \cite{2023MNRAS.526.2441B}. This prescription, based on the original EAGLE Ref sub-grid of \cite{eagle.schaye.2015} and implemented in the \textsc{SWIFT} hydrodynamics simulation code \citep{2024MNRAS.530.2378S}, produced too small or absent CCs, defined by the central entropy excess, $K_0$ \citep{entropy_profiles_pratt2010}. Overall, the entropy profiles were flat for all the 27 objects considered at $z=0$, corresponding to entropy levels in the core ($0.15\, r_{500}$) up to 1 dex too high compared to observations of like-mass objects \citep{entropy_profiles_sun2009, entropy_profiles_pratt2010}. On the other hand, recent studies of the X-GAP groups \citep{2024Galax..12...24E} highlighted several entropy profiles that are significantly flatter than previous observations.

Other simulation codes implementing cosmological accretion, feedback and cooling were also found to produce flat or very shallow entropy profiles with small or absent cool cores. We have previously shown this effect for EAGLE \citep{eagle.schaye.2015} and Cluster-EAGLE \citep{barnes_ceagle}. Other works \citep[see][and references therein]{2021Univ....7..209O} found similar results for SIMBA \citep{2019MNRAS.486.2827D} and FABLE \citep{2018MNRAS.479.5385H}. The entropy profiles in IllustrisTNG also show a plateau around cluster cores similar to that of \citetalias{Altamura2023} and return a CC fraction much lower than observations when selecting objects by $K_0$ \citep{2018MNRAS.481.1809B}. Breaking the trend, higher-resolution models that ignore metal-line cooling above a gas temperature of $10^4$~K, like ROMULUS-C \citep{2017MNRAS.470.1121T}, or model isolated halos without cosmological accretion, like \cite{2022MNRAS.515.4838N} and \cite{2022MNRAS.516.3750H}, produced a relatively long-lived cool core and a power-law-like entropy profile. Moreover, recent work by \cite{husko2024_winds} suggests that even in idealised objects, entropy profiles can be sensitive to the AGN feedback implementation: thermal isotropic schemes, as in the SWIFT-EAGLE model, tend to produce high entropy in the core, as do kinetic jets without fine-tuned jet particle velocity. However, a combination of the two schemes (\textit{hybrid} prescription, in their work) can generate stable power-law-like profiles in the $10^{14-15}$~M$_\odot$ range, while $10^{13}$~M$_\odot$ groups still produce an entropy plateau in the core.

More recent analyses of the TNG-Cluster sample \citep{nelson_tng_cluster}, a re-simulation of 352 cluster regions from the TNG300 volume, showed that the IllustrisTNG model produces CC clusters, being located at one extremum of a unimodal distribution \citep{lehle_tng_cluster} (in contrast with the bimodal $K_0$ distribution of \cite{entropy_profiles_pratt2010}), but their abundance depends strongly on the selection criteria, confirming the results of \cite{2018MNRAS.481.1809B}. Some TNG-Cluster entropy profiles also show an isentropic flat core, similarly to the cluster in \citetalias{Altamura2023}.

The formation of isentropic cores may be the outcome of the entropy-truncation mechanism, proposed by \cite{1997MNRAS.289..955K, 2000ApJ...544L...1B, 2001Natur.414..425V}, and \cite{2002ApJ...576..601V}. According to this process, the IGM gas subject to extreme cooling (condensation) or extreme heating no longer contributes to the entropy distribution, which becomes \textit{truncated} at low values of $K$ and manifests as an entropy floor in radial profiles. Specifically, the low-entropy gas subject to cooling was shown to be \textit{preferentially} heated by AGN feedback and then ejected from the system \citep{McCarthy2011}. 
In the same work, SNe feedback was shown to regulate the entropy truncation in OWLS proto-groups and proto-clusters. SN can moderate the cooling of the gas at high redshift, thus preventing the removal of low-entropy gas by star formation too early and without shifting the median entropy of the IGM. 

The sub-grid model in \cite{McCarthy2011} that included metal cooling, star formation and feedback \citep[denoted \textit{ZCool\_SF\_SN\_AGN}, as part of the OWLS suite of simulations, ][]{2010MNRAS.402.1536S} successfully reproduced X-ray properties of galaxy groups, including entropy, temperature, and density distributions, as well as their halo-mass dependence. In addition to full-physics simulations, the study presented non-radiative runs (labelled as \textit{NoCool}), where baryonic physics was switched off. By comparing thermodynamic properties of gas particles inside $r_{500}$\footnote{$r_{\Delta}$ is the radius at which the internal mean density exceeds the critical density of the Universe by a factor of $\Delta$. This approach defines $r_{500}$ for $\Delta=500$ and $r_{200}$ and for $\Delta=200$ (conventionally taken as the virial radius). $M_{\Delta}$ is defined as the total mass within a spherical overdensity of radius $r_{\Delta}$ centred in the gravitational potential minimum.} in \textit{NoCool} with their counterparts in full-physics runs, \cite{McCarthy2011} could identify the gas in ZCool\_SF\_SN\_AGN that had been ejected as a direct consequence of feedback. Reportedly, the gas ejected from the progenitors of the halo at high redshift ($z>1$) forms the high-entropy phase, while the remaining gas, albeit hot, allows the self-regulation of heating and cooling and the formation of stable cool cores.

Moreover, the increase in entropy in the hot halo gas in ZCool\_SF\_SN\_AGN is due to two factors: first, the low-entropy gas can be directly and selectively heated by the central super-massive black hole (cSMBH) and ejected from the system, allowing higher entropy gas to flow inwards to replace it; second, low-entropy gas can quickly cool and form stars, resulting in the low-entropy gas phase being further depleted. These processes have their biggest impact on low-mass progenitor halos, particularly at redshifts $z\approx 2-4$, and dictate which systems can develop a stable CC and which ones will be turned into NCC. Overall, the OWLS simulations \citep{2010MNRAS.406..822M}, as well as cosmo-OWLS \citep{2014MNRAS.441.1270L} and BAHAMAS \citep{mccarthy_bahamas}, produced steep, power-law-like entropy profiles and realistic hot gas fractions simultaneously on group scales. In the simulations of \citetalias{Altamura2023}, the entropy amplification appears to have been vastly more intense, removing most of the inner IGM gas from the low-entropy phase and inhibiting the formation of stable cool cores by $z=0$.

In this paper, we take a further step in the analysis of the group and cluster of \citetalias{Altamura2023} by examining the evolution of their properties and entropy distribution. Using time-series data, we capture the state of the group and cluster systems when the central entropy increased and characterise the processes that may have seeded the plateau in the entropy profiles. By introducing a simulation of the group at a resolution 8 times higher than EAGLE (labelled as high-res), we show that entropy plateaus persist as sub-grid physics is resolved on smaller scales. Notably, to our knowledge, our high-res group is one of the highest-resolution full-physics simulations evolved to $z=0$ of an object on group scales (the HR group in \citealt{2021MNRAS.501.4657R} was run with a similar resolution).
We divide our analysis into two sections:
\begin{itemize}
    \item First, we analyse 3D aperture properties and entropy profiles during three key phases of the evolution of the two objects. This approach is similar to the study on the ROMULUS-C cluster \citep{2017MNRAS.470.1121T, 2021MNRAS.504.3922C}.

    \item Second, we introduce the \textit{non-radiative} group and cluster simulations (see Table \ref{tab:resolutions-comparison-other-works}). By selecting matching gas particles in the non-radiative and Ref runs, we reconstruct the thermodynamic Lagrangian history of the gas in the core ($r<0.15\, r_{500}$), the entropy-plateau region ($0.15<r/r_{500}<1$), and the surrounding environment ($1<r/r_{500}<6$). The goal of this analysis is to verify the extent to which the ejection scenario described by \cite{McCarthy2011} applies to our simulations.
\end{itemize}

This work is structured as follows. In Section~\ref{sec:dataset}, we summarise the set-up and sub-grid models of the simulations used in our study. Then, in Section~\ref{sec:analysis-methods} we illustrate the analysis techniques and define the key physical quantities. Section~\ref{sec:results-evolution} contains the results with the evolution of basic properties and thermodynamic profiles, the Lagrangian histories and distributions of selected subsets of gas particles. Finally, we discuss and summarise the results in Section~\ref{sec:discussion-and-conclusions}.
Throughout this work, we assume the Planck 2018 cosmology, given by 
$\Omega_{\rm m}=0.3111$, 
$\Omega_{\rm b}=0.04897$, 
$\Omega_\Lambda=0.6889$, 
$h=0.6766$, 
$\sigma_8=0.8102$, 
$z_{\rm reion}=7.82$, 
$T_{\rm CMB}=2.7255$ K \citep{planck.2018.cosmology}.

\begin{table*}
    \setlength{\tabcolsep}{3pt}
    \begin{threeparttable}
    \centering
    \caption{Mass resolution and softening lengths for different types of simulations. The simulations introduced in \protect\citetalias{Altamura2023} are highlighted in orange, and the high-resolution group introduced in this paper is highlighted in yellow. All highlighted runs are used in our analysis. The mass of DM particles is expressed by $m_{\rm DM}$ and the initial mass of gas particles is $m_{\rm gas}$. $\epsilon_{\rm DM,c}$  and $\epsilon_{\rm gas,c}$ indicate the comoving Plummer-equivalent gravitational softening length for DM particles and gas particles respectively, $\epsilon_{\rm DM,p}$ and $\epsilon_{\rm gas,p}$ indicate the physical maximum Plummer-equivalent gravitational softening length for DM and gas particles respectively. In works using moving mesh codes, such as \protect\cite{2021MNRAS.506..488B}, $r_{\rm cell}$ indicates the cell radius.}
    \label{tab:resolutions-comparison-other-works}
    \begin{tabular}{lllccccccc}
    \toprule
    Simulation set-up                          & Model                 & Reference                    & $m_{\rm DM}$           & $m_{\rm gas}$         & $\epsilon_{\rm DM,c}$ & $\epsilon_{\rm gas,c}$ & $\epsilon_{\rm DM,p}$ & $\epsilon_{\rm gas,p}$ & Produces \\ 
                                    &                       &                               & (M$_\odot$)            & (M$_\odot$)           & (ckpc)         & (ckpc)         & (pkpc)        & (pkpc)       &cool cores?\\
    \midrule
    \rowcolor{orange!25}
    Group and cluster (low-res)     & SWIFT-EAGLE           & \protect\cite{Altamura2023}    & $7.85 \times 10^{7}$   & $1.47 \times 10^{7}$  & 6.66           & 3.80           & 2.96          & 1.69        & No \\
    \rowcolor{orange!25}
    Group and cluster (mid-res)     & SWIFT-EAGLE           & \protect\cite{Altamura2023}    & $9.82 \times 10^{6}$   & $1.83 \times 10^{6}$  & 3.33           & 1.90           & 1.48          & 0.854       & No \\
    \rowcolor{yellow!25}
    Group (high-res)                & SWIFT-EAGLE           & This work                     & $1.23 \times 10^{6}$   & $2.29 \times 10^{5}$  & 1.67           & 0.95           & 1.74          & 0.427       & No \\
    \rowcolor{yellow!25}
    Group and cluster (low-res)     & Non-radiative         & This work                     & $7.85 \times 10^{7}$   & $1.47 \times 10^{7}$  & 6.66           & 3.80           & 2.96          & 1.69        & Yes \\
    \rowcolor{yellow!25}
    Group and cluster (mid-res)     & Non-radiative         & This work                     & $9.82 \times 10^{6}$   & $1.83 \times 10^{6}$  & 3.33           & 1.90           & 1.48          & 0.854       & Yes \\
    \rowcolor{yellow!25}
    Group (high-res)                & Non-radiative         & This work                     & $1.23 \times 10^{6}$   & $2.29 \times 10^{5}$  & 1.67           & 0.95           & 1.74          & 0.427      & Yes  \\
    Two groups and cluster$^\star$  & SWIFT-EAGLE$^\dagger$ &  \protect\cite{2022MNRAS.515.4838N}   & --                     & $(1-42) \times 10^{5}$& --             & --             & --            & 0.30          & Yes \\
    Two groups and cluster$^\star$  & SWIFT-EAGLE$^\ddagger$&  \protect\cite{2022MNRAS.516.3750H}   & --                     & $(1-42) \times 10^{5}$& --             & --             & --            & 0.30          & Yes \\
    EAGLE 100 Mpc (Ref)             & EAGLE Ref             &  \protect\cite{eagle.schaye.2015}     & $9.60 \times 10^{6}$   & $1.80 \times 10^{6}$  & --             & --             & --            & 0.700      & No  \\
    EAGLE 25 Mpc (high-res)         & EAGLE Ref             &  \protect\cite{eagle.schaye.2015}     & $1.21 \times 10^{6}$   & $2.26 \times 10^{5}$  & --             & 1.33           & --            & 0.350      & No  \\
    C-EAGLE/Hydrangea               & EAGLE-AGNdT9          &  \protect\cite{barnes_ceagle}         & $9.60 \times 10^{6}$   & $1.80 \times 10^{6}$  & --             & --             & --            & 0.700      & No  \\
    TNG50                           & TNG                   &  \protect\cite{2019MNRAS.490.3234N}   & $4.50 \times 10^{5}$   & $8.50 \times 10^{4}$  & 0.288          & 0.074          & --            & --         & No  \\
    TNG300 + TNG-Cluster            & TNG                   &  \protect\cite{nelson_tng_cluster}    & $6.1 \times 10^{7}$    & $1.2 \times 10^{7}$   & 1.480          & 0.370          & --            & --         & Yes$^a$  \\
    SIMBA (m100n1024)               & SIMBA                 & \protect\cite{2019MNRAS.486.2827D}    & $9.6 \times 10^7$      & $1.82 \times 10^7$    & 0.74 & 0.74 & -- & -- & No \\
    FABLE (clusters)                & FABLE                 & \protect\cite{2018MNRAS.479.5385H}    & $8.1\times10^7$        & $9.4\times10^6$       & 4.142 & 4.142 & 4.142 & 4.142 & No \\
    ZCool\_SF\_SN\_AGN (groups)     & cosmo-OWLS            & \protect\cite{McCarthy2011}           & $5.56\times10^8$       & $1.18\times10^8$      & 10.7 & 10.7  & 2.74 & 2.74 & Yes \\
    MACSIS (390 custers)            & BAHAMAS               &  \protect\cite{barnes_macsis}         & $6.49\times10^9$       & $1.18\times10^9$      & 5.9 & 5.9 & 5.9 & 5.9 & Yes \\
    BAHAMAS (groups)                & BAHAMAS               &  \protect\cite{mccarthy_bahamas}      & $5.50\times10^9$       & $1.09\times10^9$      & 5.7 & 5.7 & 5.7 & 5.7 & Yes \\
    FLAMINGO (m9)                   & FLAMINGO              &  \protect\cite{2023MNRAS.526.4978S}   & $5.65\times10^9$       & $1.07\times10^9$      & 22.3 & 22.3 & 5.70 & 5.70 & Yes$^b$ \\
    ROMULUS-C                       & ROMULUS               &  \protect\cite{2019MNRAS.483.3336T}   & $3.40 \times 10^{5}$   & $2.10 \times 10^{5}$  & --             & --             & --                    & 0.250 & Yes \\
    Groups \textit{c10kHR}$^\star$  & Kinetic feedback      &  \protect\cite{2016MNRAS.461.1548B}   & --                     & $2.13 \times 10^{6}$  & --             & --             & --                    & 1.40  & Yes \\
    Cluster                         & Fiducial jet-AGN      &  \protect\cite{2021MNRAS.506..488B}   & $6.86 \times 10^{7}$   & $1.94 \times 10^{7}$  & --             & --             & $2.5\, r_{\rm cell}$  &  4.00 & Yes \\
    Perseus-like cluster$^{\star}$    & XMAGNET             &  \protect\cite{2025arXiv250213213G}   & --                     & --                    & --             & --             & --                    &  0.0977 & Yes \\
    \bottomrule
    \end{tabular}
    \begin{tablenotes}
      \small
      \item $^\star$Indicates full-physics simulations of isolated halos, with the highest particle-mass resolution in the inner 100 kpc and quadratically degrading resolution out to $r_{500}$, except for \protect\cite{2016MNRAS.461.1548B}, where the particle mass is fixed inside $r_{500}$, and for \protect\cite{2025arXiv250213213G}, where the maximally-refined region spans 125 kpc. The DM halo is modelled with an external potential following a Navarro-Frenk-White profile \protect\citep{nfw1997_profile}. 
      \item $^\dagger$The EAGLE-like model differs from \protect\cite{Altamura2023} in the black-hole accretion and the star-formation scheme. 
      \item $^\ddagger$The EAGLE-like model implements the changes of \protect\cite{2022MNRAS.515.4838N}, and models AGN feedback using self-consistent jets \citep{2022MNRAS.516.3750H}. The CC/NCC abundances and entropy profiles are documented in: $^a$\protect\cite{lehle_tng_cluster}; $^b$\protect\cite{2024MNRAS.533.2656B}.
    \end{tablenotes}
    \end{threeparttable}
\end{table*}

\section{Simulated group and cluster}
\label{sec:dataset}
In this analysis, we use the simulated group and cluster in \citetalias{Altamura2023}, run with the Ref and non-radiative (NR) models. This section summarises the key properties of the simulation technique and the properties of the two systems. 

\subsection{Zoom-in simulation technique and set-up}
\label{sec:dataset:technique}
The group and cluster objects are selected from a (300 Mpc)$^3$ dark-matter-only volume at $z=0$. By choice, these halos are relatively isolated systems, as they are not surrounded by any other object with a mass above 10 per cent of their mass within a $10\, r_{500}$ radius. The group has a mass $M_{500}=8.83\times 10^{12}$ M$_\odot$ at $z=0$, and was randomly selected from the low-mass sub-set of \citetalias{Altamura2023}. It is representative of small galaxy assemblies a few times more massive than the Local Group \citep[$M\approx 4\times 10^{12}$ M$_\odot$, ][]{2022ApJ...928L...5B}. Going to higher masses, the cluster has $M_{500}=2.92\times 10^{14}$ M$_\odot$ at $z=0$ and was randomly selected from the high-mass bin. The cluster is representative of a system about 3.5 times more massive than the Virgo galaxy cluster \citep[$M_{500}\approx 0.83\times 10^{14}$ M$_\odot$, ][]{2017MNRAS.469.1476S} or half the mass of the Coma galaxy cluster \citep[$M_{500}\approx 6\times 10^{14}$ M$_\odot$, ][]{2020MNRAS.497.3204M}.

To model the group and cluster, we follow the zoom-in simulation technique, which consists of refining the numerical resolution inside a defined volume of interest and degrading the resolution elsewhere \citep{1993ApJ...412..455K, 1997MNRAS.286..865T}. By concentrating the resolution elements in the region of interest, the zoom-in method is particularly efficient at modelling individual objects with moderate computational cost. The zoom-in initial conditions are generated in three steps:
\begin{enumerate}
    \item The high-resolution region is chosen to be a $6\, r_{500}$ sphere, centred on the potential minimum at $z=0$. The positions of the DM particles in this volume are then tracked back to the initial conditions at $z=127$ using their unique IDs.
    \item The Lagrangian volume enclosing the selected particles at $z=127$ is masked, in preparation for the high-resolution particle load. In some instances, the Lagrangian domain presents concave sub-regions, such as gaps or holes. 
    We extrude the mask boundaries until the main concavities are filled to minimise the chances of contamination from low-resolution DM-filled gaps. This technique is referred to as \textit{topological closure} and is described in Chapter~3 of \cite{2023altamuraPhD}. 
    With topological closure, the number of high-resolution DM particles increases by $1-5$ per cent to fill the extra padding around the Lagrangian volume. Despite a slightly longer run-time, the additional refinement leads to a nearly contamination-free volume out to $6\, r_{500}$. 
    \item The displacement and velocity fields are computed using \textsc{Panphasia} \citep{2013MNRAS.434.2094J} and \textsc{IC\_gen}, a \textsc{Fortran} code that implements the method of \cite{2010MNRAS.403.1859J} to generate realisations of primordial Gaussian random fields in a multi-scale setting\footnote{The phase descriptor containing the unique seed of the white noise field used for the parent simulation and the zoom simulations is $\rm [Panph1,L18,(74412,22732,260484),S3,CH1799108544,EAGLE-XL\_L0300\_VOL1]$.}.
\end{enumerate}

The initial conditions for the group and cluster are produced at three resolutions, labelled `low', `mid' and `high', summarised in Table~\ref{tab:resolutions-comparison-other-works} (highlighted). The mid resolution is the same as for the EAGLE L100N1504 volume from \citet{eagle.schaye.2015}, characterised by dark matter particles with mass $m_{\rm DM}=9.82 \times 10^{6}~{\rm M}_\odot$ and initial gas particle mass $m_{\rm gas}=1.83 \times 10^{6}~{\rm M}_\odot$. With a mass resolution 8 times lower than EAGLE, the low-resolution setup has 8 times larger (initial) particle masses. Finally, the high-resolution setup has 8 times smaller particle masses, making it 8 times higher resolution than the EAGLE L100N1504 volume. Table~\ref{tab:resolutions-comparison-other-works} also reports the corresponding comoving and physical Plummer-equivalent gravitational softening length for DM and gas particles associated with the three resolutions. 

We run dark-matter-only simulations of the two objects at all resolutions to check for contaminating low-resolution (i.e. boundary) dark matter particles used to define the tidal field around the high-resolution region within the $6\, r_{500}$ clean radius. At $z=0$, the contaminating mass within $6\, r_{500}$ was at most 0.1 per cent, with no boundary particles within $r_{500}$.

\subsection{The SWIFT-EAGLE reference model}
\label{sec:dataset:ref}
We summarise the features of the SWIFT-EAGLE Ref model, referring to \citetalias{Altamura2023} for a full discussion. As in the original EAGLE Ref model of \cite{eagle.schaye.2015}, the thermodynamics of the ISM is governed by a pressure floor \citep{2008MNRAS.383.1210S}. The hydrodynamics of the gas is modelled using the smoothed-particle hydrodynamics (SPH) approach, particularly suited for describing systems with a vast density dynamic range, such as galaxy formation environments. The simulations were run with \textsc{SWIFT} configured with the SPHENIX \citep{sphenix_borrow2022} hydrodynamic scheme. SPHENIX implements the `traditional' density-energy method, in addition to artificial conduction and artificial viscosity.

The gas cooling process is governed by the \cite{2020MNRAS.497.4857P} cooling tables. This prescription accounts for a redshift-dependent UV/X-ray background from \cite{2020MNRAS.493.1614F}, self-shielding for the cool ISM, presence of dust, interstellar radiation field and cosmic rays via the Kennicutt–Schmidt relation \citep{1998ApJ...498..541K}. When cold and dense gas reaches the equation of state of the pressure floor, it is converted into stars. Star formation is modelled as a sub-grid process. It uses a Kennicutt–Schmidt-like pressure law \citep{1959ApJ...129..243S, 1998ApJ...498..541K, 2008MNRAS.383.1210S} and a metallicity-dependent number density threshold \citep{2004ApJ...609..667S, 2015MNRAS.450.1937C}, above which gas turns into stars. 

Each star particle represents a simple stellar population (SSP), which contributes to the chemical enrichment of the surrounding gas via the sub-grid treatment of three stellar wind channels \citep{2009MNRAS.399..574W}: AGB and massive stars, core-collapse supernovae (SNII) and type-Ia supernovae \citep[SNIa,][]{eagle.schaye.2015}. Assuming a \cite{2003PASP..115..763C} initial mass function (IMF) populated by stars between 0.1-100~M$_\odot$, we consider all massive stars above 8~M$_\odot$ to end their life with SNII episodes. In each time-step, the expected number of SNII events, each releasing $10^{51}$ erg of energy, is computed. Notably, the probabilistic approach of this calculation makes our SN feedback implementation stochastic. Finally, the total energy is then injected thermally into a small number of neighbouring gas particles, as prescribed by \cite{2012MNRAS.426..140D}. During this step, the temperature of gas particles involved in the feedback is raised by $\Delta T_{\rm SN}=10^{7.5}$ K. The gas particles to heat are chosen to be closest to the SSP, as prescribed by the \textit{minimum-distance} energy distribution scheme (\citetalias{Altamura2023}).

The feedback from AGN is powered by SMBHs, seeded at $z\leq19$ in groups above a friends-of-friends (FoF) mass of $M_{\rm FoF}=10^{10}$~M$_\odot$. With an initial mass set to $M_{\rm seed}=10^{4}$~M$_\odot$, (supermassive) black holes (abbreviated as BHs) are positioned at the gas density maximum. Subsequently, they can grow by accreting surrounding gas or merging with other BHs.

Using the spherically symmetric Bondi–Hoyle–Lyttleton model \citep{1939PCPS...35..405H, 1944MNRAS.104..273B}, the mass accretion rate is expressed as 
\begin{equation}
    \label{eq:bh_accretion}
    \dot{m}_{\rm BH} = \alpha ~ \frac{4 \pi G^2~m_{\rm BH}^2~\rho_{\rm gas}}{\left(c_{\rm s}^2 + v_{\rm gas}^2\right)^{3/2}},
\end{equation}
where $m_{\rm BH}$ is the BH subgrid mass, $c_{\rm s}$ is the gas sound speed, $v_{\rm gas}$ the bulk velocity of the gas in the kernel relative to the BH, and $\rho_{\rm gas}$ the gas density evaluated at the BH position. 
The SWIFT-EAGLE BH accretion model does not require boosting \citep{2009MNRAS.398...53B} because of the sufficiently high gas density around the BHs. Therefore, the boost factor is $\alpha=1$. We do not apply an Eddington limiter to the mass accretion rate to allow for realistic quasar-mode accretion, and, similarly to the AGN feedback in the EAGLE-like model in \cite{2022MNRAS.515.4838N}, we do not include the \cite{2015MNRAS.454.1038R} angular momentum limiter. Finally, we measure $\dot{m}_{\rm BH}$ in units of the \cite{1926ics..book.....E} accretion rate, which scales linearly with BH mass: $\dot{m}_{\rm Edd}\propto m_{\rm BH}$.

Our AGN feedback mechanism is deterministic and uses an energy reservoir to trigger the energy injection. A fraction $\epsilon_{\rm r}=0.1$ \citep[radiative efficiency from][]{1973A&A....24..337S} of the mass accreted by the BH is converted into energy, and a fraction $\epsilon_{\rm f}=0.1$ of that energy is then accumulated in the reservoir. When the reservoir contains enough energy to raise the temperature of a designated gas neighbour by $\Delta T_{\rm AGN}$ \citep{2009MNRAS.398...53B}, an AGN feedback event occurs. In the Ref model, the temperature of the target gas particle is raised by $\Delta T_{\rm AGN}=10^{8.5}$ K, fixed for all AGN feedback events. The single gas particle to heat per feedback event is chosen to be closest to the BH, in analogy with the SN feedback scheme. In summary, this AGN feedback scheme is purely thermal (no kinetic energy is directly transferred to the gas) and deterministic. To compensate for the unresolved dynamical friction, BHs need to be artificially re-positioned towards the gravitational potential minimum \citep[\texttt{Default} method of][]{bahe_2021_bh_repositioning}. This algorithm prevents the BHs from drifting away from the centre of galaxies into low-density environments, where they would grow unnaturally slowly and generate little feedback activity. Unlike the repositioning method of \citetalias{Altamura2023}, the target potential minimum is obtained excluding the contribution of the BH to the local gravitational potential. This change improves the stability of the algorithms by reducing the risk of high-mass BHs drifting towards self-induced gravitational potentials. With the BHs stably located in high-density environments, we expect their final mass to be higher and the AGN feedback stronger than in \citetalias{Altamura2023}.

\subsection{The non-radiative model}
\label{sec:dataset:other-models}
To directly assess the impact of baryonic processes, we rerun our group and cluster with photo-heating, but switching off radiative cooling. Star formation, BH seeding, and feedback are not included either. This configuration, describing a model that only includes gravity, gas hydrodynamics, and a photo-ionising UV/X-ray background \citep{2020MNRAS.493.1614F}, is referred to as a non-radiative (NR) model. As in \cite{McCarthy2011}, we use this model of the group and cluster to track identical particles to that in Ref to probe the effects of baryonic processes on the IGM (see Section \ref{sec:lagrangian}).

\section{Analysis methods}
\label{sec:analysis-methods}

Our study employs two types of analysis. The first type, used in Section \ref{sec:results-evolution}, defines the global properties of the objects within $r_{500}$ and the radial profiles, sampled at different snapshots throughout the simulation; we describe these properties in Section \ref{sec:analysis-methods:properties}. The second type is based on selecting identical sets of particles using their unique IDs, as required to construct the Lagrangian thermal histories presented in Section \ref{sec:lagrangian}; we describe this technique in Section \ref{sec:analysis-methods:selections}.

\subsection{IGM properties}
\label{sec:analysis-methods:properties}
We define the gas mass, $M_{\rm gas}$, by adding the mass of the gas particles within $r_{500}$
\begin{equation}
    M_{\rm gas} = \sum_{i:r<r_{500}} m_{g,i},
\end{equation}
where $r$ is the Euclidean distance from the potential minimum and $m_{g,i}$ is the mass of the $i^{\rm ith}$ gas particle. This ensemble of particles is also subject to a temperature selection criterion depending on whether the hot or cold gas mass is computed. The stellar mass $M_\star$ is obtained similarly, but replacing the gas-particle mass with the star particle mass $m_{\star,i}$.

The entropy calculation follows \citetalias{Altamura2023}. Firstly, we obtain the number density of free electrons accounting for the chemical abundances from the chemical elements tracked by the sub-grid model. Assuming fully ionised gas, we define the total free-electron fraction for each hot ($T>10^5$ K) gas particle as
\begin{equation}
    X_{\rm e} \equiv \frac{n_{\rm e}}{n_{\rm H}} = \frac{m_{\rm H}}{f_{\rm H}}~\sum_\epsilon Z_\epsilon \frac{f_\epsilon}{m_\epsilon}
\end{equation}
and the ion fraction $X_i$
\begin{equation}
    X_{\rm i} \equiv \frac{n_{\rm i}}{n_{\rm H}} = \frac{m_{\rm H}}{f_{\rm H}}~\sum_\epsilon \frac{f_\epsilon}{m_\epsilon},
\end{equation}
which can be combined to compute the electron number density
\begin{equation}
    n_{\rm e} = \frac{X_{\rm e}}{X_{\rm e} + X_{\rm i}} \frac{\rho_g}{\mu~m_{\rm H}} = \rho_g ~ \sum_\epsilon Z_\epsilon \frac{f_\epsilon}{m_\epsilon},
\end{equation}
where $f_\epsilon$ is the gas-particle mass fraction for chemical element $\epsilon = \left\{{\rm H, He, C, N, O, Ne, Mg, Si, S, Ca, Fe}\right\}$, $m_\epsilon$ the associated atomic mass and $Z_\epsilon$ the atomic number. $f_{\rm H}$ and $m_{\rm H}$ are the hydrogen mass fraction and atomic mass respectively. Then, $\rho_g$ is the SPH density of the gas particle and $\mu$ its mean atomic weight, computed by combining the contribution from the chemical species as
\begin{equation}
    \mu = \left[ m_{\rm H} \sum_\epsilon \frac{f_\epsilon}{m_\epsilon} ~ (Z_\epsilon + 1) \right]^{-1}.
\end{equation}
For our results, we present entropy, mass-weighted temperature and density profiles of the simulated groups and clusters. To compute the radial profiles, we consider 50 spherical shells centred on the halo's centre of potential with a logarithmically increasing radius, spanning $(0.01-2.5)\, r_{500}$. We then sum the contributions of the particles within each radial bin. 
For the $i^{\rm th}$ shell, we compute the density profile as
\begin{equation}
    \rho_{g,i} = \frac{\sum_j m_j}{V_i},
\end{equation}
where the sum is the total gas mass in shell $i$, divided by the volume $V_i$ of the shell. We also define the mass-weighted temperature $T_{{\rm MW}, i}$ as
\begin{equation}
    T_{{\rm MW}, i} = \frac{\sum_j m_j T_j}{\sum_j m_j}.
\end{equation}
The entropy profiles are then computed via the mass-weighted temperature profiles and the density profiles, as described in \cite{vikhlinin_2006_profile_slope}
\begin{equation}
    K(r)=\frac{\mathrm{k_B}T(r)}{n_{\rm e}(r)^{2/3}}.
\end{equation}
The entropy profiles are normalised to their self-similar values, appropriate for an atmosphere in hydrostatic equilibrium. The normalisation is obtained in three steps. Firstly, we define the critical density of the universe as
\begin{equation}
    \rho_{\rm crit}(z) = E^2(z) \frac{3 H_0^2}{8 \pi G},
\end{equation}
where $E^2(z)\equiv H^2(z) / H_0^2 =\Omega_{\rm m}(1+z)^3 + \Omega_\Lambda$. To select X-ray emitting gas, we only consider gas particles above a temperature of $10^5$ K. Secondly, we compute the characteristic temperature at $r_{500}$,
\begin{equation}
    k_\mathrm{B}T_{500}=\frac{G \bar{\mu} M_{500} m_{\rm H}}{2\,r_{500}},
\end{equation}
where $\bar{\mu} = 0.5954$ is the mean atomic weight for an ionised gas with primordial ($X = 0.76$, $Z = 0$) composition. Finally, using the characteristic temperature and density ($500\, \rho_{\rm crit}$), we establish the characteristic entropy:
\begin{equation}
    K_{500}=\frac{k_\mathrm{B}T_{500}}{\left[500 f_{\rm bary} \rho_{\rm crit}~/(\overline{\mu}_{\rm e} m_{\rm H})\right]^{2/3}},
\end{equation}
where $\overline{\mu}_{\rm e} = 1.14$ is the mean atomic weight per free electron and $f_{\rm bary}=0.157$ is the universal baryon fraction obtained by \cite{planck.2018.cosmology}.

\subsection{Matched particle ensembles}
\label{sec:analysis-methods:selections}

Selecting particles using only a spatial mask, such as the $r<r_{500}$ criterion for the cluster properties, does not guarantee that the set contains the same particles between different snapshots. In fact, within this mask, gas particles are regarded as \textit{indistinguishable}. Crucially, baryonic particles are expected to be displaced at almost every time-step, following the equations of motion, e.g. SPH particles \citep{sphenix_borrow2022}, or repositioning algorithms, i.e. SMBHs \citep{bahe_2021_bh_repositioning}. 

To characterise the thermodynamic history of the particles that describe the IGM today, it is necessary to select \textit{identical} particles unequivocally, which can be achieved only if they are \textit{distinguishable}. In SWIFT,  particles can be distinguished by their unique ID, a long-integer value assigned to each particle at the start of the simulation and never altered. Using particle IDs, our selection strategy is described as follows. Firstly, we generate a list containing the unique IDs of particles that satisfy given criteria at $z=0$; we denote this set as the \textit{master list}. Secondly, we search for particles in a different snapshot whose ID appears in the master list. This technique was introduced in \cite{McCarthy2011} to construct the Lagrangian history of sets of unique particles and track their thermodynamic state through redshift. Similar methods employing massless tracers were successful in tracking fluid elements in mesh-based simulations to study the entropy core in galaxy clusters via velocity-field tracers \citep{2011MNRAS.410..461V} or Monte Carlo tracers \citep{2013MNRAS.435.1426G}. In all our SPH simulations, we adopt the method by \cite{McCarthy2011} with additional selection criteria, as described below.

Throughout this work, particles selected at $z=0$ are required to be in the hot, X-ray-emitting phase, i.e. above $10^5$ K. For instance, in the case of a selection by particle ID on the present-day Ref snapshot we denote this selection as $\mathrm{ID_{Ref}}(z=0, T>10^5~\mathrm{K})$. Since we always select particles by ID, we can abbreviate this notation as $\mathcal{A}_\mathrm{Ref}=\mathrm{Ref}(z=0)$. Following this logic, the particles from any snapshot are represented as $\mathcal{B}_\mathrm{Ref}=\mathrm{Ref}(z\geq0)$. Note that no temperature cut is imposed on the $z>0$ snapshots since we only require the gas `parcels' to be in the X-ray-emitting phase at the present day. The particle tracking is performed by searching the particle IDs in $\mathcal{B}$ that also appear in $\mathcal{A}$, expressed as follows:
\begin{equation}
    \mathcal{B}_\mathrm{Ref} \in \mathcal{A}_\mathrm{Ref} \equiv \mathrm{Ref}(z\geq0) \in \mathrm{Ref}(z=0, T>10^5~\mathrm{K}).
\end{equation}

Provided that the initial conditions are identical, the comparison between snapshots is not restricted to the same simulation but can be performed between different runs, here Ref and NR. In our study, we consider four combinations, described as follows.
\begin{enumerate}
    \item $\mathcal{B}_{\rm Ref} \in \mathcal{A}_{\rm Ref}$. This choice selects the particles at $z=0$ in Ref and tracks them back in time in the Ref model. 
    \item $\mathcal{B}_{\rm NR} \in \mathcal{A}_{\rm NR}$ operated as the case above but for the NR model. 
    \item $\mathcal{B}_{\rm NR} \in \mathcal{A}_{\rm Ref}$. This choice selects particles in Ref at $z=0$ but tracks them throughout redshift in NR.
    \item $\mathcal{B}_{\rm Ref} \in \mathcal{A}_{\rm NR}$. Now, we select particles in NR but then track them throughout redshift in Ref. This procedure is analogous to the previous one, but swapping the model indices: NR$\, \leftrightarrow\, $Ref.
\end{enumerate}

By construction, the master list in Ref excludes particles that formed stars or were swallowed by BHs. Therefore, the gas at higher redshift that is destined to be converted into other particle types or energy does not feature in the Lagrangian histories built with our prescription. The same is done when matching particles between different models: the master list for case (iv) also excludes NR particles that do not have a Ref counterpart at $z>0$ due to star formation or BH accretion, i.e. the list only includes particles whose ID exists in both lists. This restriction is not necessary in case (iii) because no gas particles are lost in the NR scenario, where sub-grid physics is switched off. To simplify the ID tracking, we exclude the gas particles that have undergone splitting, implemented in \textsc{SWIFT} to prevent particles from becoming very massive when they have absorbed a large amount of stellar mass loss. We do not expect this approximation to affect the Lagrangian history since the gas particles involved in splitting are $<0.1$ per cent of the ensemble for the simulation with the most splitting events (high-res group).

Since we aim to investigate the shape of the entropy profile (and not just the entropy level), we split the $z=0$ selection into three independent spherical regions, as follows: 

\begin{enumerate}
    \item \textbf{Core.} This master list includes particles that end up hot in the core, as defined by the additional spatial mask: $T>10^5~\mathrm{K} \land r < 0.15\, r_{500} \land z=0$.

    \item \textbf{Shell.} This master list includes hot particles within the IGM, but located outside the core:  $T>10^5~\mathrm{K} \land 0.15 < r/r_{500} < 1 \land z=0$. The shell surrounding the core appears to contain hot gas with approximately similar entropy, which we identify as an entropy plateau in radial profiles.

    \item \textbf{Outskirts.} Lastly, we select hot particles that end up outside the group or cluster, i.e. $r > r_{500}$. We restrict the outer boundary to the \textit{clean} refined region in the zoom-in simulation, which is $6\, r_{500}$ at $z=0$. Thus, the constraints for this master list are: $T>10^5~\mathrm{K} \land 1 < r/r_{500} < 6 \land z=0$.
\end{enumerate}


\section{Redshift evolution}
\label{sec:results-evolution}

We present results from the evolution history of the group and cluster in four parts: first, we show the evolution of useful properties and, based on them, identify three key phases; second, we qualitatively illustrate the shift from cold- to hot-gas-dominated IGM; third, we focus on the evolution of the baryon content (quantitatively); fourth, we link the above results to the scaled entropy profiles.

\subsection{Basic properties}
\label{sec:results-evolution:basic}

\begin{figure*}
    \centering
    \includegraphics[width=\textwidth]{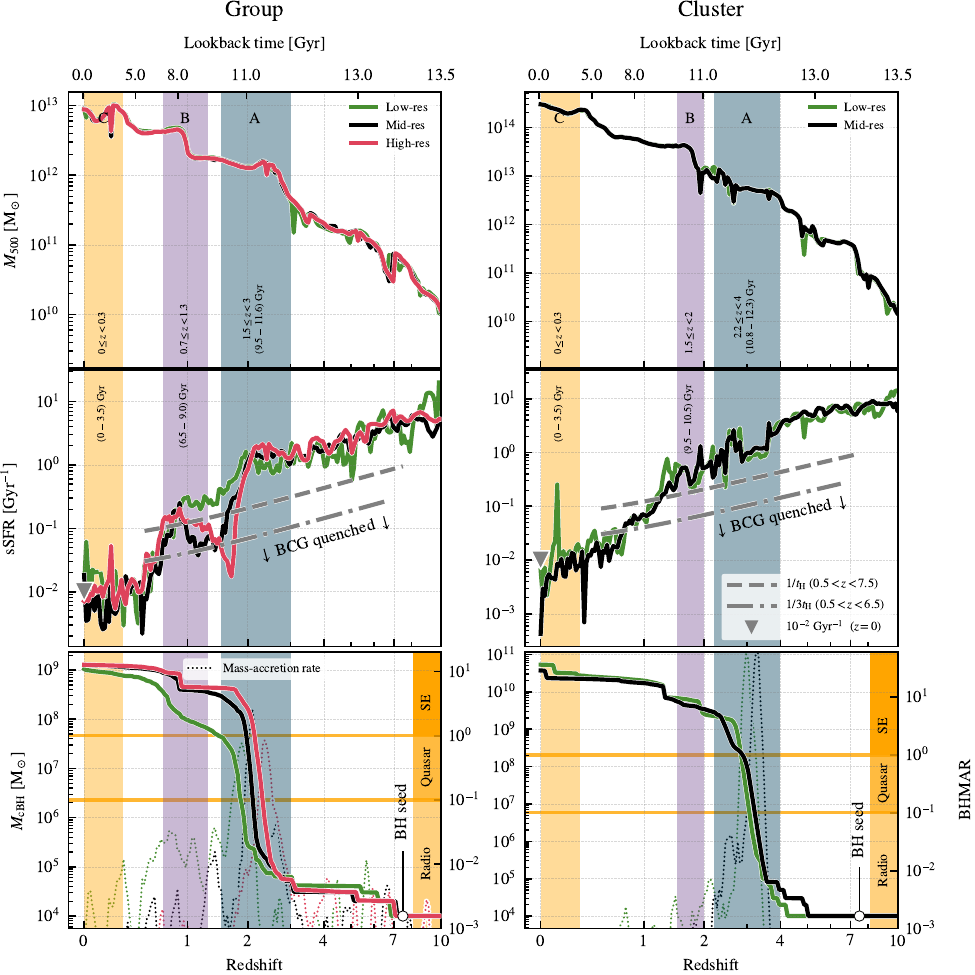}
    \caption{Evolution of the mass ($M_{500}$, top row), the specific star-formation rate (sSFR, middle row) in a physical 50 kpc aperture, and the mass of the central black hole (cBH, bottom row) with redshift, $z$, and lookback time (top axis). \textit{Left}: properties of the group at low (green), mid (black) and high (red) resolution. \textit{Right}: properties of the cluster at low (green) and mid (black) resolution. The three phases are indicated with coloured bands, with the intervals reported in terms of redshift (top row) and lookback time (middle row). The intervals are fixed for both objects. We indicate sSFR$=10^{-2}$ Gyr$^{-1}$, below which we consider the brightest central galaxy (BCG) quenched, with a grey triangle at $z=0$ \citep{2022MNRAS.512.1052P}. We also show two quenching threshold definitions at high redshift: $1/t_{\rm H}(z)$, where $t_{\rm H}(z)$ is the redshift-dependent Hubble time, for $0.5<z<7.5$ (dashed grey), and $1/3t_{\rm H}(z)$ for $0.5<z<6.5$ \citep[dot-dashed grey,][]{2019MNRAS.487.5416T, 2024ApJ...961..163B}. In the bottom row, we highlight the seed mass of the central BH, $M_{\rm cBH}=10^4$ M$_\odot$. The dotted lines are the BH mass accretion rate (BHMAR) in units of the Eddington rate, averaged over ten rolling snapshots using a Savitzky-Golay filter with a degree-3 polynomial basis \citep{savitzky1964smoothing}. The BHMAR scale is shown on the right. The horizontal yellow lines are the threshold BHMAR that defines the AGN to be in radio mode, $\mathrm{BHMAR} < 10^{-3}\, \dot{m}_{\rm Edd}$, quasar mode, $10^{-3} \leq \mathrm{BHMAR} / \dot{m}_{\rm Edd} < 1$, and super-Eddington (SE) mode, $\mathrm{BHMAR} \geq \dot{m}_{\rm Edd}$.}
    \label{fig:properties-evolution-1}
\end{figure*}

\begin{table}
    \centering
    \caption{Summary of the three phases of the evolution of the group and cluster considered in this work. We report the redshift ($z$) and lookback time ($t$, in Gyr) intervals with a description of the event or physical state characterising each evolutionary phase. The phase description is colour-coded as in the rest of the document; the $z$ and $t$ ranges are highlighted in alternating colours for readability.}
    \label{tab:time-phases}
    \begin{tabular}{ccrcc}
    \toprule
    Phase & Description & & \textbf{Group} & \textbf{Cluster} \\
    \midrule
    \multirow{2}{*}{\color{colorbanda}{\textbf{A}}} & \multirow{2}{*}{\color{colorbanda}{Quasar mode}}  & \cellcolor{lightgray!30} $z$ & \cellcolor{lightgray!30} $1.5-3.0$   & \cellcolor{lightgray!30} $2.2-4.0$     \\
                                                    &                                                   & $t$ & $9.5-11.6$  & $10.8-12.3$   \\
    \multirow{2}{*}{\color{colorbandb}{\textbf{B}}} & \multirow{2}{*}{\color{colorbandb}{Merger}}       & \cellcolor{lightgray!30} $z$ & \cellcolor{lightgray!30} $0.7-1.3$   & \cellcolor{lightgray!30} $1.5-2.0$     \\
                                                    &                                                   & $t$ & $6.5-9.0$   & $9.5-10.5$    \\
    \multirow{2}{*}{\color{colorbandc}{\textbf{C}}} & \color{colorbandc}{Present state}                 & \cellcolor{lightgray!30} $z$ & \cellcolor{lightgray!30} $0-0.3$     & \cellcolor{lightgray!30} $0-0.3$       \\
                                                    & \color{colorbandc}{(quenched)}                    & $t$ & $0-3.5$     & $0-3.5$       \\
    \bottomrule
    \end{tabular}
\end{table}

In Fig.~\ref{fig:properties-evolution-1}, we show the evolution of the halo mass ($M_{500}$), the specific star-formation rate (sSFR) in Gyr$^{-1}$, the BH mass (in ${\rm M}_\odot$), and BH mass accretion rate (BHMAR) in units of the Eddington rate $\dot{m}_{\rm Edd}$ computed for the central BH only. These quantities are defined as follows. $M_{500}$ is the total mass within $r_{500}$, as discussed previously. The sSFR is computed using star particles within a 50 kpc spherical aperture and using stars formed in the 50 Myr before the snapshot. At $z=0$ (local Universe), we call the BCG quenched if its sSFR is below the quenched threshold of $10^{-2}$ Gyr$^{-1}$, indicated by a grey triangle \citep[][from SDSS DR7 galaxies with stellar mass between $10^{9}-10^{12}$~M$_\odot$ and halo mass $>10^{11}$~M$_\odot$]{2022MNRAS.512.1052P}. At high redshift ($z>0.5$), we use two redshift-dependent quenched thresholds, defined as $1/t_{\rm H}(z)$ (dashed grey) and $1/3t_{\rm H}(z)$ (dot-dashed grey). These were shown to be effective in classifying active and quenched galaxies in the IllustrisTNG \citep{2019MNRAS.487.5416T} and JWST-CEERS samples \citep{2024ApJ...961..163B} up to $z=7.5$ and $z=6.5$ respectively \citep[for JWST-CEERS observations, higher-redshift bins contain too few quenched galaxies to estimate a threshold reliably, see also][]{2023ApJ...944..108B}. The Eddington accretion rate  \citep{1926ics..book.....E} used to scale the BHMAR is defined for a spherically symmetric matter distribution as
\begin{equation}
    \dot{m}_{\rm Edd} = \frac{4 \pi G m_{\rm P}}{\epsilon_{\rm r}~c~\sigma_{\rm T}} ~ m_{\rm BH} \approx 2.218~{\rm M_\odot~yr^{-1}}~\frac{\epsilon_{\rm r}}{0.1}~\left( \frac{m_{\rm BH}}{10^8~{\rm M_\odot}} \right).
\end{equation}
We track the progenitor of the main halo at $z=0$ by taking the object with the largest $M_{\rm FoF}$ in the high-resolution volume at $z>0$. Here, we use halo centres computed with the VELOCIraptor structure-finding code configured with the 6-dimensional FoF algorithm \citep{2019PASA...36...21E}. We found this method sufficient for tracking the central object's progenitors over most of the redshift range, as shown qualitatively by the $M_{500}$ evolution in Fig.~\ref{fig:properties-evolution-1}. Occasionally, the mass jumps to lower values when a secondary halo near the central one becomes the most massive for a short time. These inaccuracies do not affect the redshift-evolution results. In Section~\ref{sec:results-evolution:entropy-profiles}, we manually excluded the affected snapshots within the redshift bands from the analysis. The results of Section~\ref{sec:lagrangian} are unaffected because they only use unique particle IDs selected at $z=0$.

We identified three evolutionary phases, summarised in Table \ref{tab:time-phases} and described below. 
\begin{itemize}

    \item \textbf{Phase A (quasar mode)} spans $z=1.5-3.0~(2.2-4.0)$ for the group (cluster) and captures the peak of AGN activity. During this period, the proto-group is a galaxy with a median mass $M_{500}=1.38 \times 10^{12}$~M$_\odot$ in rapid growth. The BCG experiences intense star formation, which sharply declines below the $1/3t_{\rm H}(z)$ quenched threshold by the end of the phase as the central BH’s mass accretion rate approaches $\dot{m}_{\rm Edd}$. The proto-cluster's mass increases rapidly to that of a fully formed galaxy (median $M_{500}=5.31 \times 10^{12}$~M$_\odot$). During this phase, the central BH also grows rapidly, and its AGN feedback will later lead to the shutdown of star formation in the BCG.

    \item \textbf{Phase B} spans $z=0.7-1.3~(1.5-2.0)$ for the group (cluster) and captures \textbf{merger} events which further increase $M_{500}$. In both objects, the sSFR of the BCG fluctuates between $10^{-1}-10^{0}$~Gyr$^{-1}$; the central BH’s mass increases.

    \item \textbf{Phase C} represents the \textbf{quenched} state of the systems near the present time ($0 \leq z < 0.3$). The \textit{group} shows a decrease in $M_{500}$, followed by an increase due to the infall of a substructure, correlating with an increase in sSFR. The BHMAR remains just below $\approx 10^{-3}\, \dot{m}_{\rm Edd}$. The \textit{cluster} also experiences a mass accretion event at low redshift. By the end of phase C, both the group and cluster central BHs are in \textit{radio mode}, defined by BHMAR~$< 10^{-3}\, \dot{m}_{\rm Edd}$ \citep[e.g.][]{2009ApJ...698.1550H}.

\end{itemize}

The metrics for the group and the cluster at different resolutions are broadly consistent throughout the simulation, indicating good numerical convergence. However, we find that the sSFR of the BCG in the group at low-res is higher than for the other two resolutions by up to 1 dex between $z=1-2$ (phase B). This is because the central BH is less massive compared to mid- and high-res during that time, and therefore the feedback it produces is less effective in quenching star formation.

The brief dips in $M_{500}$ are artefacts of the \textsc{VELOCIraptor} structure-finding code, which picked the wrong halo as the central during the merger of similar-mass objects. This effect could be corrected using merger-tree information; however, this operation is unnecessary for our study, as the artefacts do not affect our conclusions. The gaps in sSFR at redshift 10 or higher indicate periods where no BCG-associated stars formed within a 50 Myr window.

We summarise the insights from Fig.~\ref{fig:properties-evolution-1} as follows. The growth of the group and cluster progenitors via mergers favours the BH growth. Consequently, this intensifies the AGN activity, leading to a drop in sSFR below the quenching threshold in (part of) phase B. This result supports previous work that correlates the growth of BHs with merger activity \citep{2015ApJ...799..178K} and the BH self-regulation mechanism \citep{2012MNRAS.425L..66M, 2022MNRAS.515.4838N}. Moreover, these events are more noticeable in the group progenitor than the cluster progenitor.

\subsection{Transition from the cold to hot IGM}
\label{sec:dataset:objects}

\begin{figure*}
    \includegraphics[width=2\columnwidth]{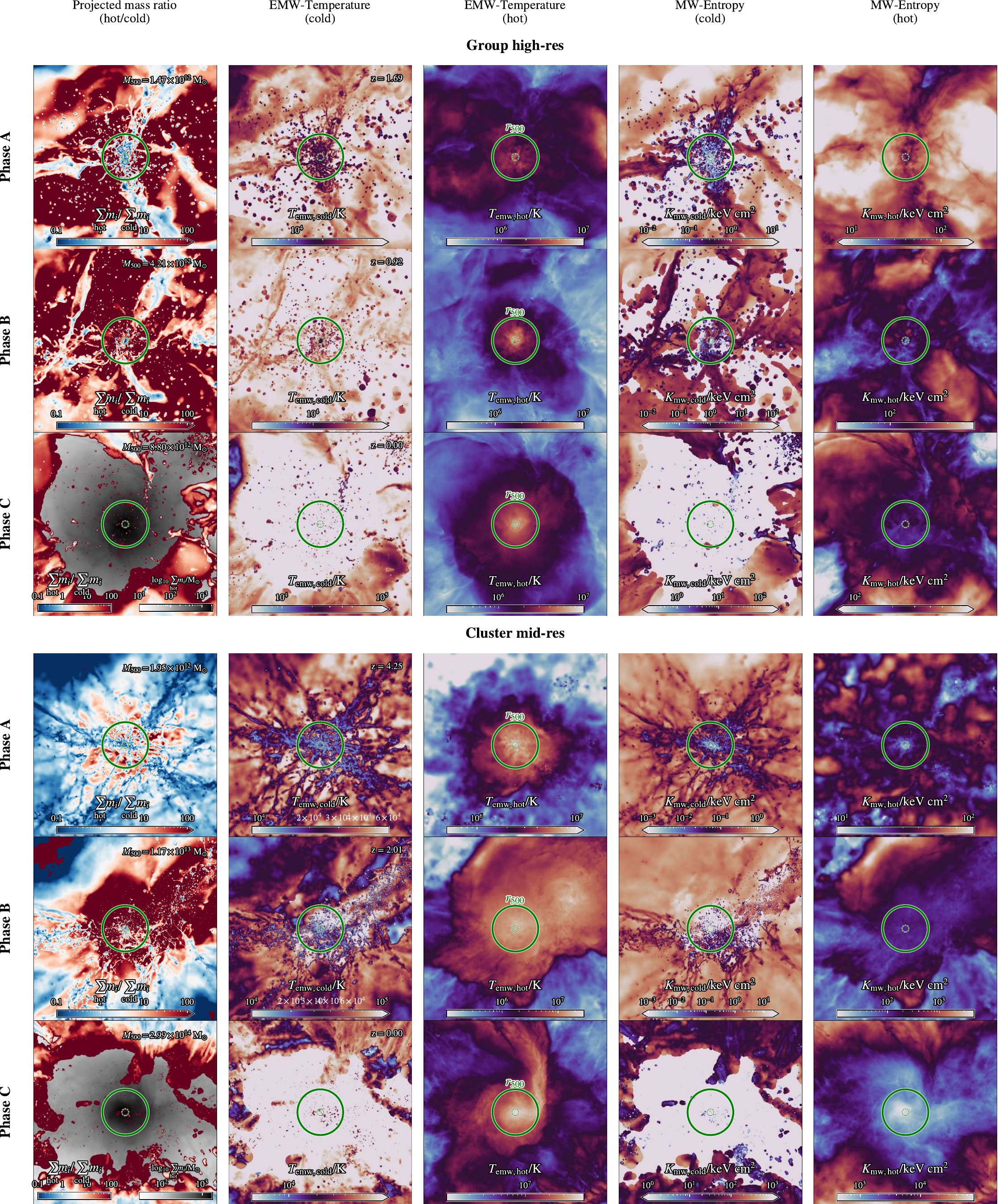}
    \caption{Projected maps of the group (top three rows) and the cluster (bottom three rows), run with the Ref model at high- and mid-res, respectively. From top to bottom, the rows correspond to phase A (quasar mode), B (merger) and C (quenched). The five columns show the following projected quantities, from left to right: (i) ratio of the projected \textit{hot} gas mass to projected \textit{cold} gas mass, as defined in Eq.~\ref{eq:project-mass-ratio}; (ii) the emission-measure-weighted temperature of the cold gas, followed by (iii) the same but for hot gas; finally, (iv) the mass-weighted projected entropy of the cold gas and (v) the same but for hot gas. The projected mass ratio is masked in solid red where all gas is hot; white areas indicate no particles matching the temperature selection in other maps. For phase C, we overlay the projected mass ratio without cold gas (in grey) with the map of the hot gas to visualise the underlying gas distribution. The projections along the $z$-axis are constructed from a spatial cube of (physical) volume $(4\, r_{500})^3$ around the centre of the objects. $r_{500}$ and the core radius, $0.15\, r_{500}$, are indicated by green solid and dashed circles respectively.}
    \label{fig:accretion-maps}
\end{figure*}

The evolution of the IGM gas temperature is key to understanding the entropy distribution and the baryon content. In Fig.~\ref{fig:accretion-maps}, we illustrate temperature and entropy maps for the high-res group and the mid-res cluster at three redshifts within each phase, as indicated to the left of the panels. The columns show the following quantities, from left to right. The \textbf{projected mass ratio}, $R_{i}$, is evaluated by summing the contributions of particles along the line of sight (LoS), smoothed by the Wendland-C2 kernel $W$ \citep{wendland1995piecewise} implemented in \textsc{SWIFTsimIO} \citep{Borrow2020}. Thus, the contributions of the $j^{\rm th}$ particle to the $i^{\rm th}$ pixel is computed as
\begin{equation}
    R_i=\frac{\sum_{j\,\in\,{\rm LoS}}m_{i,\rm hot}\,W_{i,j}(h_j)}{\sum_{j\,\in\,{\rm LoS}}m_{i,\rm cold}\,W_{i,j}(h_j)},
    \label{eq:project-mass-ratio}
\end{equation}
where $m_{i,\rm hot}$ are the gas particle masses with temperature $T\geq10^5$~K, $m_{i,\rm cold}$ those with $T<10^5$~K, and $h_j$ is the SPH smoothing length of the $j^{\rm th}$ particle. The ratio in Eq.~\ref{eq:project-mass-ratio} is shortened to $\sum_{\rm hot} m_i / \sum_{\rm cold} m_i$ in Fig.~\ref{fig:accretion-maps}. The maps show the projected mass ratio with blue shades indicating a LoS with more cold gas than hot, and red shades indicating the opposite. In the centre, areas without any cold gas along the LoS are shown in grey. 

The two A$\rightarrow$B$\rightarrow$C map sequences show the evolution of the group and cluster's IGM from being mainly cold (phase A), to having similar hot and cold gas fractions during merger events (phase B, the second row of Fig.~\ref{fig:properties-evolution-2}), and finally to a hot-gas phase (C). For the projected mass ratio in the latter phase, we overlay the map of the hot gas mass (in grey shades) to render the underlying diffuse hot gas distribution within and around $r_{500}$. All quantities are given in the physical reference frame at the corresponding redshift.

The maps in columns 2-3 and 4-5 show the emission-measure-weighted (EMW) \textbf{temperature} and the mass-weighted (MW) \textbf{entropy}, respectively; each pair breaks down the IGM gas into the cold (2 and 4), and hot (3 and 5) phases. The EMW temperature for the $i^{\rm th}$ pixel is defined as 
\begin{equation}
    T_{i,\rm emw}=\sum_{j\,\in\,{\rm LoS}}m_i\rho_i\,T_i\,W_{i,j}(h_j) \Big/ \sum_{j\,\in\,{\rm LoS}}m_i\rho_i\,W_{i,j}(h_j),
\end{equation}
where the emission measure is $\propto m_i\rho_i$ and $\rho$ is the SPH gas density. The MW entropy is defined similarly but without the $\rho$-weights:
\begin{equation}
    K_{i,\rm mw}=\sum_{j\,\in\,{\rm LoS}}m_i\,K_i\,W_{i,j}(h_j) \Big/ \sum_{j\,\in\,{\rm LoS}}m_i\,W_{i,j}(h_j),
\end{equation}
with $K_i=k_{\rm B}T_i/n_{e,i}^{2/3}$.
Most IGM gas is cold in phase A and starts with temperatures and entropies lower than the self-similar values. During phase B, when the to-be group has the size of a galaxy and the to-be cluster has the size of a group, the amount of hot gas increases in the central region and cold, low-entropy gas streams form in the surroundings. By $z=0$ (phase C), hardly any cold gas can be found within $r_{500}$; the entropy maps show a uniform distribution on scales of $\approx r_{500}$ for the group and scales of $\approx 0.3\, r_{500}$ for the cluster, as \citetalias{Altamura2023} found previously.

\subsection{Baryon content}
\label{sec:results-evolution:baryons}

\begin{figure*}
    \centering
    \includegraphics[width=\textwidth]{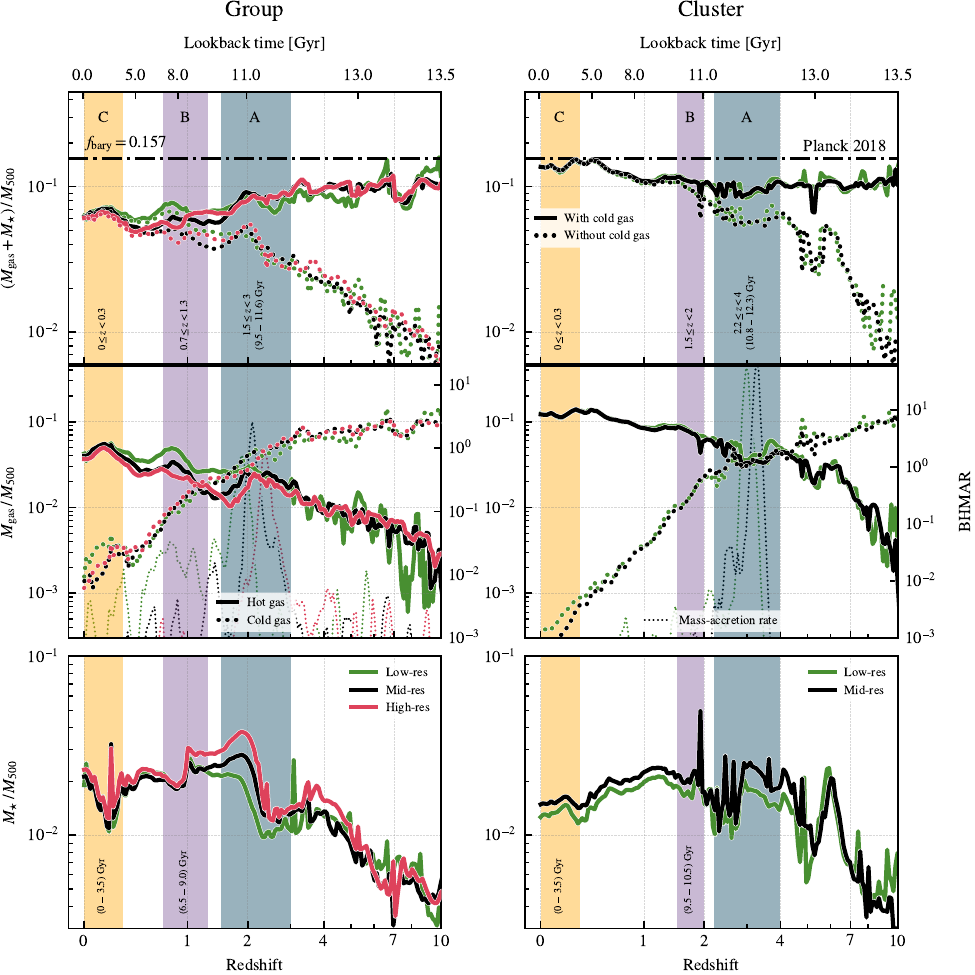}
    \caption{From top to bottom, we show the redshift/time evolution of the baryon fraction [$(M_{\rm gas} + M_{\star}) / M_{500}$], the hot gas fraction ($M_{\rm gas} / M_{500}$), the star fraction ($M_{\star} / M_{500}$) and the BCG mass fraction ($M_{\rm BCG} / M_{\star}$) within $r_{500}$. In the top panels, solid lines indicate the baryon fraction accounting for \textit{all} gas (cold and hot), while the dotted lines only include hot gas. The dash-dotted line shows the Universal baryon fraction $f_{\rm bary}=0.157$ for the Planck 2018 cosmology \protect\citep{planck.2018.cosmology}. In the middle panels, the solid lines show the fraction of \textit{hot} gas and the (thick) dotted lines show the fraction of \textit{cold} gas; the thin dotted lines represent the BHMAR. The stellar component of the BCG is defined by a spherical aperture with a fixed physical radius of 50 kpc. The $x$-axes and the layout are as in Fig. \ref{fig:properties-evolution-1}.}
    \label{fig:properties-evolution-2}
\end{figure*}

The baryon, gas and star fractions are displayed in Fig.~\ref{fig:properties-evolution-2} with the same layout as in Fig.~\ref{fig:properties-evolution-1}. First, we define these additional metrics and then describe the main features of their evolutionary profiles. 

The gas mass $M_{\rm gas}$ is computed by summing the mass of all gas particles within $r_{500}$. $M_{\rm gas}$ is equivalent to the sum of the \textit{hot} gas mass $M_{\rm gas,hot}$, which accounts for gas above $10^5$~K, and the \textit{cold} gas mass $M_{\rm gas,cold}$, which accounts for gas below $10^5$~K. The hot and cold gas fractions are obtained by dividing their respective masses by $M_{500}$. In the middle row of Fig.~\ref{fig:properties-evolution-2}, the gas fraction $f_{\rm gas} = M_{\rm gas}/M_{500}$ is split into hot (solid lines) and cold (dotted lines). The stellar mass $M_{\star}$ is the sum of the mass of all stellar particles inside $r_{500}$. From this definition, the star fraction is expressed as $f_{\star} = M_{\star} / M_{500}$. Finally, the baryon fraction is the sum of the gas and star fractions: $f_{\rm gas} + f_{\star} \equiv (M_{\rm gas} + M_{\star}) / M_{500}$. In the top row of Fig. \ref{fig:properties-evolution-2}, the solid lines show the baryon fraction including all gas (hot and cold), while the dotted lines exclude cold gas.

Focusing on the top row of Fig.~\ref{fig:properties-evolution-2}, the (total) baryon fraction of both objects starts close to the Universal value of $f_{\rm bary}=0.157$ \citep{planck.2018.cosmology} at $z\approx10$. This result is consistent with little SN and AGN feedback at such early times (see also the BHMAR in Fig. \ref{fig:properties-evolution-1}). However, the late-time evolution for $z<2$ shows the outcomes of two competing effects: the deepening of the gravitational potential well (hence accretion) and the intensified feedback activity. In the group, the total baryon fraction gradually decreases because of the gas ejected by feedback. On the other hand, the cluster's gravitational potential is strong enough to hold onto the gas despite the feedback, causing an \textit{increase} in the total baryon fraction. This result indicates that feedback has a stronger effect on groups than in clusters, in line with previous numerical \citep{2010MNRAS.406..822M, McCarthy2011, 2011MNRAS.415.1549G} and observational \citep{xxl.baryons.akino2022, 2023MNRAS.520.6001P} studies, and physical intuition.

The impact of AGN feedback on the IGM raises the question of whether a different variation of the feedback model could lead to a net loss of baryons for the cluster too. In our previous investigation of this particular cluster \citepalias{Altamura2023}, we did not find evidence that a different feedback energy-injection scheme or moderately more explosive AGN feedback could have ejected more baryons in the form of hot gas. Tests with a higher AGN heating temperature ($\Delta T_{\rm AGN}=10^9$ K) have shown little impact on the baryon fraction at $z=0$. However, increasing $\Delta T_{\rm AGN}$ did lead to higher entropy levels in the core, thereby exacerbating the core entropy excess relative to the self-similar expectation. The total time-integrated energy available for AGN feedback was kept constant in these experiments.

From early times until the beginning of phase A, cold gas is the primary component of the baryon content in both systems: the cold gas fractions are high, $f_{\rm gas,cold}\approx0.1$ and the hot gas fractions are low, $f_{\rm gas,hot}\approx0.01$. This is also reflected in the baryon fraction with and without cold gas (top row of Fig. \ref{fig:properties-evolution-2}). During this phase, the central black holes have not started their non-linear growth phase yet, and the AGN activity remain low. However, by $z\approx4$ the sSFR begins to decrease, suggesting that the abundant cold gas feeds star formation, and SN feedback regulates the IGM thermodynamics in halos of mass scales $\sim 10^{12}$ M$_\odot$, i.e. small galaxies. This result is consistent with the equilibrium mechanisms modelled in \citet{2014MNRAS.444.2071D}.

When the halo mass reaches $\sim 10^{12}$ M$_\odot$ (galaxy scale, phase A), the gas heats up, and the SN feedback is no longer effective \citep[e.g.][]{2017MNRAS.465...32B}. Instead, the gas can now efficiently feed the cSMBH and the BHMAR reaches its peak, which clearly coincides with the decrease of $f_{\rm gas,cold}$ below $f_{\rm gas,hot}$ in Fig.~\ref{fig:properties-evolution-2} (middle row). By the end of phase B, the cSMBH grows further and AGN feedback becomes the main process regulating the IGM thermodynamics and shutting down star formation. We find this effect in both objects; however, it is most apparent in the group, where a significant dip in sSFR happens earlier. Moreover, the cSMBH appears to reach the peak of its growth rate when the halo has mass $\approx 10^{12}$ M$_\odot$, independent of the redshift, in agreement with the original EAGLE simulations \citep[see Fig.~6 of][]{2017MNRAS.465...32B}. Also in line with the same study, the cSMBHs hosted by our simulated halos have slowed down their growth rate by the end of the merger phase, once the system surpassed the `transition mass'.

The resulting IGM state (phase C) features quenched BCGs and small cold gas fractions. For both objects, we find that higher-resolution simulations produce more stars, except for the low-res group at low redshift with a nearly converged $f_{\star}$. This feature can be observed in most phases of the $f_{\star}$ evolution and is in line with Fig.~5 of \citetalias{Altamura2023}.

\subsection{Entropy profiles}
\label{sec:results-evolution:entropy-profiles}

\begin{figure*}
    \centering
    \includegraphics[width=\textwidth]{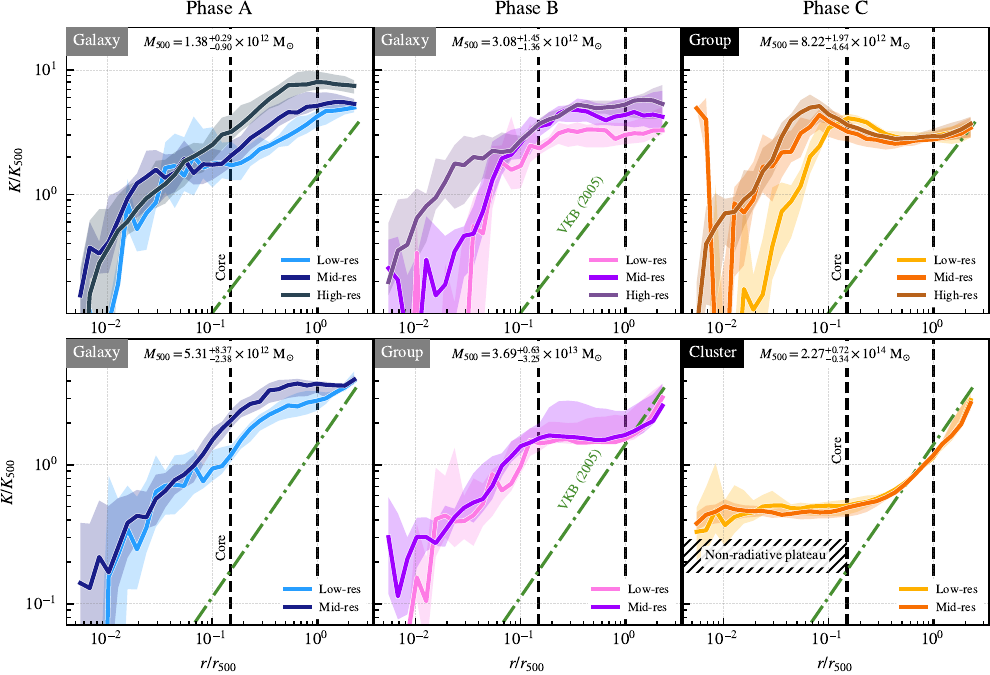}
    \caption{Scaled entropy profiles for the group (top) and the cluster (bottom) during the three evolutionary phases in Table~\ref{tab:time-phases}. The grey labels indicate the mass scale of the halo in each phase, and the black labels indicate the $z=0$ mass scale of the objects. The solid lines represent the median entropy profiles in the respective redshift interval, and the shaded region is the first and third quartile levels. From left to right: \textit{Phase A} covers the high-redshift (quasar-mode) evolution, \textit{Phase B} the intermediate redshift during mergers, and \textit{Phase C} captures the low-redshift thermodynamic state ($z < 0.3$). The colour coding for the three phases is the same as in Fig.~\ref{fig:properties-evolution-1}. In each panel, the runs at different resolutions are shown with varying saturation and luminance in the same colour-space region. For the group, we show the low-, mid-, and high-res entropy profiles, while for the cluster we only include the low- and mid-res profiles. The profiles are plotted out to $2.5\, r_{500}$. The green dash-dotted line is the self-similar non-radiative entropy baseline \citep[VKB,][]{vkb_2005}, and the vertical dashed lines are guidelines for the core radius ($0.15\, r_{500}$, as indicated) and $r_{500}$. The $M_{500}$ mass, laid out as $\left(\rm{median}\right)^{+(\rm{max}-\rm{median})}_{-(\rm{median}-\rm{min})}$ for each phase is shown at the top of each panel. The hatched region in the cluster's entropy profiles indicates the entropy level of the plateau in the core of the corresponding NR simulations (see Appendix~\ref{app:nr-profiles}). Here, the Ref level is higher than in NR, indicating a clear entropy excess.}
    \label{fig:entropy-profile-scaled}
\end{figure*}

Now, we study the entropy profiles of the objects during each of the three evolutionary phases defined in Section \ref{sec:results-evolution:basic}. For each phase, we compute the median profile, together with the first and third quartiles. In NR conditions, the group and cluster develop clear power-law-like entropy profiles, behaving as expected. These profiles are reported in Appendix \ref{app:nr-profiles}. For the Ref model, we present the group and cluster profiles in Fig.~\ref{fig:entropy-profile-scaled}. To compute the entropy, we only include the highly-ionised gas in the X-ray emitting phase, defined by the temperature cut $T>10^5$ K. The entropy profiles are shown down to 1 per cent of $r_{500}$. In addition, we truncate the entropy profiles where the number of particles within the spherical shell falls below 50 if this radius is larger than $0.01\, r_{500}$.  

In Fig.~\ref{fig:entropy-profile-scaled}, the group entropy profiles are on the top row, while the cluster profiles are on the bottom row. The panels are arranged in three columns, with the respective evolutionary phase indicated at the top. In each panel, we show the entropy profiles at different resolutions for a given object during a fixed redshift interval. The solid lines show the median profiles, and the bands represent the extent between the first and third quartiles. Darker colours correspond to higher resolution and lighter colours to lower resolution. We outline two key features of these results: the shape of the profiles and the dependence on resolution.

\begin{figure*}
    \centering
    \includegraphics[width=\textwidth]{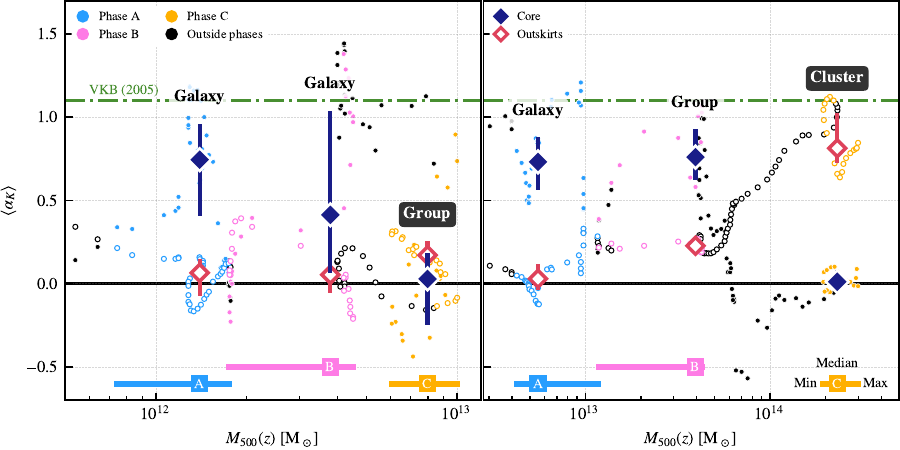}
    \caption{Evolution of the logarithmic gradient of the entropy profile $\alpha_K$ with halo growth, parametrised as $M_{500}(z)$, for the group at high-res (left) and the cluster at mid-res (right). The average gradient $\langle\alpha_K\rangle$ in the core is shown as filled markers and for the outskirts as empty markers. The small circles, one per simulation snapshot, are coloured in blue, pink and yellow, corresponding to the evolutionary phases A, B and C, respectively. Snapshots recorded outside the redshift bounds of the phases are shown in black. For each phase, the larger diamond markers show the median gradient and the first and third quartiles. The dash-dotted line is the gradient of the non-radiative gravitational baseline \citep[VKB,][]{vkb_2005}, as in Fig.~\ref{fig:entropy-profile-scaled}. At the bottom, the squares indicate each phase's median, minimum and maximum fractional halo mass. Note that the group experiences a small decrease in $M_{500}$ at the end of phase C, resulting in $x$-values slightly above $M_{500}(z=0)$.}
    \label{fig:entropy-profile-slope}
\end{figure*}

Next, we track the evolution of the logarithmic gradient of the entropy profile, defined as
\begin{equation}
    \alpha_K = \frac{d\log(K)}{d\log(r)} = \frac{d\log(K/K_{500})}{d\log(r/r_{500})}.
\end{equation}
We sample the average local gradient $\langle\alpha_K\rangle$ in two radial regions: the core, $0.05\leq r/r_{500}<0.15$, and the outskirts, $0.5\leq r/r_{500}<2$. The slope is obtained from a least-squares linear fit of the entropy profile in logarithmic space, using the data in the corresponding radial range. The two slopes, labelled $\langle\alpha_{K,\rm core}\rangle$ and $\langle\alpha_{K,\rm outskirts}\rangle$, are computed for each redshift snapshot and plotted as a function of the scaled halo mass $M_{500}(z)/M_{500}(z=0)$ in Fig.~\ref{fig:entropy-profile-slope}. Normalising $M_{500}(z)$ by its present-day value shows the group and cluster's growth within the same $x$-axis range and, hence, facilitates the comparison of the $\langle\alpha_K\rangle$ evolution. In the plot, filled markers represent $\langle\alpha_{K,\rm core}\rangle$ and empty markers $\langle\alpha_{K,\rm outskirts}\rangle$; these are coloured by the corresponding evolutionary phase, or in black if the snapshot's redshift does not correspond a particular phase. To better visualise the evolution of $\langle\alpha_K\rangle$ with noisy single-snapshot data, we smoothed the values by computing a moving average over 7 adjacent snapshots in phase A, 3 in B and 2 in C. The diminishing levels of smoothing are required by the higher statistical noise in the profiles at high redshift than at low redshift, especially in the core region.

\subsubsection{Profile shape and gradient at different mass scales}
\label{sec:results-evolution:entropy-profiles:shape}
During phase A, the to-be group is hosted by a galaxy-sized halo with $M_{500}$ just above $10^{12}$ M$_\odot$. The entropy profile is clearly power-law-like between $r_{500}$ and the core radius ($0.15\, r_{500}$). The entropy at $r_{500}$, $K(r=r_{500})$, is around an order of magnitude higher than the self-similar gravitational baseline shown in green \citep[VKB,][]{vkb_2005}, suggesting a departure from self-similarity which was also found in other simulations of galaxy-sized objects \citep[e.g.][]{2014ApJ...783L..10G}.

As the halo grows into a (larger) galaxy-sized one through mergers ($M_{500}\approx3\times10^{12}$ M$_\odot$, phase B), the entropy profile flattens at the virial radius; however, the core entropy remains largely unchanged relative to $K_{500}$. Interestingly, a similar shape appears in the profile of the other galaxy (to-be cluster) at $M_{500}\approx5\times10^{12}$ M$_\odot$ in phase A. The logarithmic entropy gradient in the two galaxies' cores and outskirts confirms this result. In Fig.~\ref{fig:entropy-profile-slope}, we compare the `galaxy' $\alpha_K$ values in the left and the right panels, corresponding to phases B and A, respectively. In the outskirts, we find $\alpha_K\approx0$ in both cases; in the core, the $\alpha_K$ values are also within the first and third quartile levels, despite the left panel showing more variability. Examining the latter case closely, the scatter is large because phase B includes $\alpha_{K, \mathrm{core}}$ samples with low values at the beginning, before the merger, and high values towards the end, after the merger, when the halo grows to the size of a large galaxy. This consideration indicates that the high $\alpha_{K, \mathrm{core}}$ values should be considered when comparing to the galaxy in the right panel, thus strengthening the similarity.

The two galaxy-sized objects show that the entropy profile at the virial radius becomes flat regardless of a merger event happening: when the halo mass reaches $3-5\times10^{12}$~M$_\odot$, the group series already recorded a merger event in phase B, but the cluster series, in phase A, has not. 

On group scales ($0.8-4\times10^{13}$~M$_\odot$ for our objects), the entropy profile is the flattest at $r_{500}$, but in the centre we still find a power law. Finally, on cluster scales ($\approx2\times10^{14}$ M$_\odot$), the gravitational power-law profile emerges at large radii, while the profile in the core flattens to a level $\approx0.2$ dex above the NR plateau.

In summary, Figs.~\ref{fig:entropy-profile-slope} and \ref{fig:entropy-profile-scaled} show that the entropy plateau appears to emerge (i) at the virial radius when the halo reaches a characteristic mass scale of $\sim 10^{12}$ M$_\odot$ and (ii) in the core at cluster scales. Since this characteristic mass scale is similar to that of the transition from SN- to AGN-feedback-driven self-regulation of the IGM \citep{2017MNRAS.465...32B}, it is natural to ask whether the plateau emerges due to processes occurring before AGN feedback or around the same time as AGN activity starts.

By definition, both objects record the peak in SMBH activity in phase A. During this time, the to-be cluster also reaches a state of `cold-hot gas equality' ($f_{\rm gas, cold}\approx f_{\rm gas, hot}$, see Fig.~\ref{fig:properties-evolution-2}) at the characteristic mass. However, in the to-be group, this occurs with a delay: the state with $f_{\rm gas, cold}\approx f_{\rm gas, hot}$ at the characteristic mass is recorded during phase B when the entropy distribution at the virial radius is flattened. Therefore, the formation of the plateau at large radii in group-mass halos appears to be due to processes that become important when the IGM becomes dominated by hot gas.

\subsubsection{Numerical convergence}
\label{sec:results-evolution:entropy-profiles:resolution}

Overall, most metrics tracked in Figs.~\ref{fig:properties-evolution-1} and \ref{fig:properties-evolution-2} show good numerical convergence at the three resolutions. On the other hand, the numerical resolution has a small impact on the scaled entropy profiles of the group in Fig.~\ref{fig:entropy-profile-scaled}. In phase A, high-res simulations produced entropy levels that are 0.3 dex higher than mid-res between the core and the virial radius. In phases B and C, the high-res profiles are higher in the core, while the entropy outside the core is well converged. This effect is due to the removal of hot gas that cools to form stars. We find that star formation is enhanced in the high-res group compared to the low- and mid-res simulations, as shown in the bottom left panel of Fig.~\ref{fig:properties-evolution-2}. 

Even in the group's case, the discrepancy between resolutions is small and does not affect our results qualitatively.

\section{Lagrangian particle histories}
\label{sec:lagrangian}

\begin{figure*}
    \includegraphics[width=2\columnwidth]{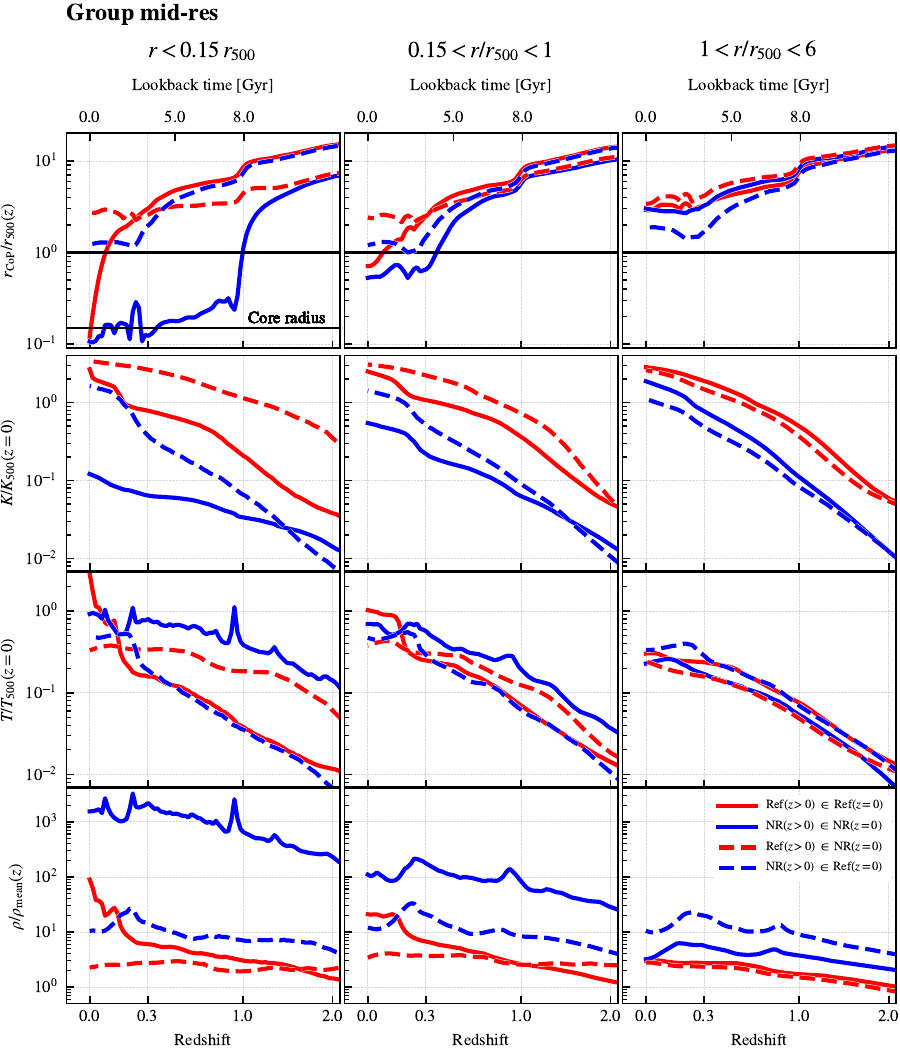}
    \caption{\textit{From top to bottom:} Lagrangian history of the median scaled halo-centric distance ($r_{\rm CoP}$), entropy ($K$), temperature ($T$), and density ($\rho$) of gas particles in the group at mid-res. \textit{From left to right:} the particles are identified in three spherical regions defined at $z=0$: the core ($r<0.15\, r_{500}$, left), the rest of the intragroup medium ($0.15 <r/r_{500}<1$, centre), and outside $r_{500}$ (right), as indicated at the top of the panels. These radii are also indicated in the overview maps of Fig.~\ref{fig:accretion-maps}. The particles in each spherical region are selected in three ways: 
    (i) gas particles in Ref that end up hot and inside the region in Ref at $z=0$ (red solid); 
    (ii) gas particles in NR that end up inside the region in NR at $z=0$ (solid blue); 
    (iii) gas particles in NR that end up in the region in Ref at $z=0$ (dashed blue); (iv) gas particles in Ref that end up in the region in NR at $z=0$ (dashed red). 
    The entropy and temperatures are normalised to the self-similar value at $z=0$; the density is normalised to the mean matter density at each redshift, $\rho_{\rm mean}(z)=\Omega_{\rm m}\, \rho_{\rm crit}(z)$; the radius is normalised to $r_{500}(z)$ and computed at each redshift. 
    The horizontal black lines indicate $r_{500}$ and the core radius, $0.15\, r_{500}$.}
    \label{fig:medians_tracked_group}
\end{figure*}

\begin{figure*}
	\includegraphics[width=2\columnwidth]{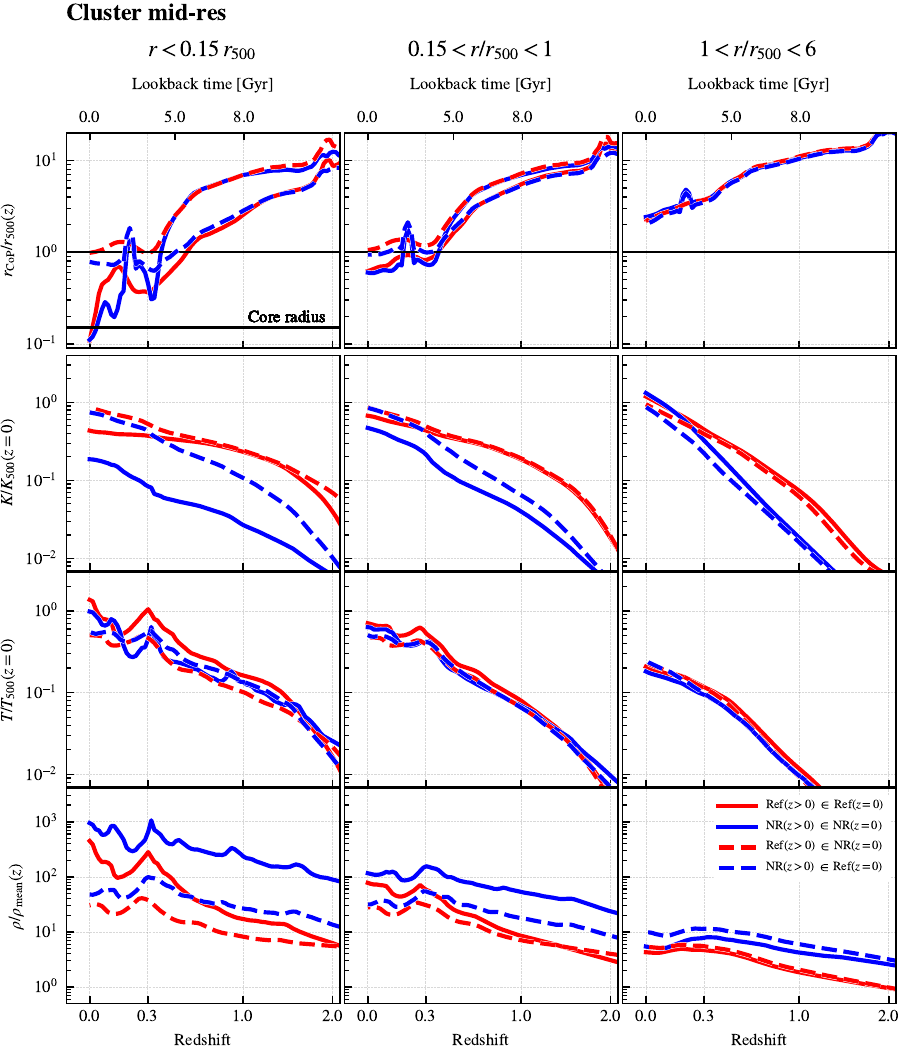}
    \caption{As in Fig.~\ref{fig:medians_tracked_group}, but for the cluster at mid-res.}
    \label{fig:medians_tracked_cluster}
\end{figure*}

One compelling hypothesis for the excess entropy, relative to the self-similar expectation, observed in our simulated objects is the hot gas ejection scenario. In this picture, initially cold gas is heated directly by SN feedback (before phase A) and then by AGNs. As its temperature increases, the gas is ejected from the halo, leaving behind a higher-entropy medium. To test this hypothesis, we match individual gas particles between our radiative (Ref) and non-radiative (NR) simulations following the procedure outlined in \citet{McCarthy2011} (see Section~\ref{sec:analysis-methods:selections}). This matching allows us to directly compare the thermodynamic histories of identical particles under different physical assumptions.

Figs.~\ref{fig:medians_tracked_group} and \ref{fig:medians_tracked_cluster} illustrate the median Lagrangian histories of key thermodynamic quantities for gas particles in the group and the cluster, respectively, at mid-res. In each figure, the rows (from top to bottom) show the evolution of the (median) scaled distance from the potential minimum $r/r_{500}(z)$, entropy $K/K_{500}(z=0)$, temperature $T/T_{500}(z=0)$, and density $\rho/\rho_{\rm mean}(z)$. Here, the self-similar values are computed at $z=0$; the density and radial distance are normalised to the mean density of the universe, $\rho_{\rm mean}(z)=\Omega_{\rm m}\, \rho_{\rm crit}(z)$, and $r_{500}(z)$ respectively.

The panels are arranged by the region in which the gas is located at $z=0$: the core ($r<0.15\, r_{500}$, left column), the intermediate intra-group/-cluster region ($0.15<r/r_{500}<1$, centre), and the outskirts ($1<r/r_{500}<6$, right).

Each panel distinguishes gas samples using a consistent colour and line code:
\begin{itemize}
    \item \textbf{Red solid:} gas particles selected in Ref that end up hot and inside the region at $z=0$, tracked in the Ref simulation. From Section~\ref{sec:analysis-methods:selections}, this particle subset is defined as $\mathrm{Ref}(z>0) \in \mathrm{Ref}(z=0)$.
    \item \textbf{Blue solid:} gas particles selected in NR that end up inside the region in NR at $z=0$ tracked in NR: $\mathrm{NR}(z>0) \in \mathrm{NR}(z=0)$.
    \item \textbf{Red dashed:} the same NR-selected particles, but tracked in the Ref simulation (restricted to gas particles that are not converted into stars or have cooled below the star formation threshold). This selection is defined by $\mathrm{Ref}(z>0) \in \mathrm{NR}(z=0)$.
    \item \textbf{Blue dashed:} gas particles selected in Ref that end up hot and inside the region at $z=0$, tracked in the NR simulation: $\mathrm{NR}(z>0) \in \mathrm{Ref}(z=0)$. This sample addresses the question `\textit{what would the history of the Ref-selected gas have been in the absence of feedback or radiative cooling?}'
\end{itemize}

Let us first consider the evolution of gas particles in the group, the system with the largest entropy excess as seen in Fig.~\ref{fig:entropy-profile-scaled}. The key plots to focus on are the halo-centric distance, $r_{\rm CoP}$ (first row), and the entropy, $K$ (second row).

Comparing the NR-selected gas ($\mathrm{NR}(z>0) \in \mathrm{NR}(z=0)$, blue solid) with the same gas tracked in Ref ($\mathrm{Ref}(z>0) \in \mathrm{NR}(z=0)$, red dashed), we find that the latter remains at several core radii away from the centre of potential. The two histories begin to diverge between $z=2$ and $1$ as a consequence of feedback processes active at high $z$. The same curves in the entropy and temperature panels (second and third row) show that this ejected gas has indeed been heated and remains at large radii, avoiding cooling down and forming stars. Moreover, the red dashed entropy is much higher than the blue solid due to their density (bottom row) being much lower; hence, the gas accreting process must be very different in Ref than in NR.

The blue dashed sample ($\mathrm{NR}(z>0) \in \mathrm{Ref}(z=0)$) has, for the most part, higher entropy than the blue solid ($\mathrm{NR}(z>0) \in \mathrm{NR}(z=0)$), which was selected in NR. Hence, the gas that replaces the ejected gas in Ref already had higher entropy \textit{initially} in NR, so that the final entropy distribution of the gas in Ref is a mixture of the initial selection (the gas required to replace the ejecta), and the sample whose entropy changed as a result of losing gas from cooling and star formation in Ref, possibly causing the red solid ($\mathrm{Ref}(z>0) \in \mathrm{Ref}(z=0)$) and blue dashed lines not to overlap perfectly. We note that comparing blue solid with red dashed ($\mathrm{Ref}(z>0) \in \mathrm{NR}(z=0)$) is not straightforward since the red dashed line represents a sample of gas particles that survived as hot gas until $z=0$ in Ref, as previously highlighted. Therefore, the red dashed sample may reflect a biased subset in NR upon matching particles. While this result is robust and useful to understand the Lagrangian entropy history, future work should further characterise the impact of this bias on the thermodynamic properties of the particle ensembles.

Now, we assess the red solid ($\mathrm{Ref}(z>0) \in \mathrm{Ref}(z=0)$) and blue dashed ($\mathrm{NR}(z>0) \in \mathrm{Ref}(z=0)$) samples; this comparison is cleaner compared to the previous one, because all the gas particles in Ref at $z=0$ can be located in NR. The Ref-selected gas ($\mathrm{Ref}(z>0) \in \mathrm{Ref}(z=0)$, red solid) which ends up in the group's core ($r<0.15\, r_{500}$), is outside the halo in NR (blue dashed) at $z=0$. The red solid and blue dashed samples have comparable median entropy despite the gas particles being located at different radii, thus proving that the hot halo gas in NR was ejected, and the Ref gas is new material that replaced it. By contrast, the difference between the blue solid and red dashed ($\mathrm{Ref}(z>0) \in \mathrm{NR}(z=0)$) curves is much greater than between the red solid and blue dashed, suggesting that the gas that has taken the place of the ejected material has not been heated as much as the ejected gas itself. In summary, much of the excess entropy in the Ref group does originate from the movement of particles to replace those that have cooled or were ejected. Upon closer inspection, the entropy of the Ref-selected gas is slightly higher than that of the NR-selected gas ($\mathrm{NR}(z>0) \in \mathrm{NR}(z=0)$, blue solid), suggesting an entropy excess for the Ref hot gas that does at least partly originate from the selection of the particles.

We also note that the entropy histories of the red solid ($\mathrm{Ref}(z>0) \in \mathrm{Ref}(z=0)$) and blue dashed ($\mathrm{NR}(z>0) \in \mathrm{Ref}(z=0)$) samples differ at $z>0.3$, with the blue dashed line increasing faster than the red solid. On the other hand, these samples have very similar histories in terms of temperature and halo-centric distance (for $z>0.3$), implying that the entropy excess is induced by a deficit in density, as shown in the third row. Near $z=0$, the self-similarly scaled temperature, density and entropy of the Ref-tracked particles (red solid) increase, while they remain roughly constant (or show an inverse trend) for the NR-tracked sample (blue dashed). This result, together with a rapidly decreasing halo-centric distance, suggests that radiative cooling in Ref drives a cooling flow which increases the gas density, and causes it to fall into the potential well. Once in the core (horizontal line), the gas compresses and undergoes shock-heating. A trace of this process can be observed in the $z=0$ projected-mass-ratio map in Fig.~\ref{fig:accretion-maps}, where a filament in the outskirts of the group is captured before sinking and collapsing.

In the intermediate shell ($0.15<r/r_{500}<1$) -- where the entropy plateau is most prominent -- the thermodynamic evolution ($T, \rho, K$) is similar to the core. At $z>0.3$, the entropy excess between the red solid and blue dashed samples is larger than the present day.

In the outskirts of the group ($1<r/r_{500}<6$), the median histories of NR-tracked particles selected in NR ($\mathrm{NR}(z>0) \in \mathrm{NR}(z=0)$, blue solid) and Ref ($\mathrm{NR}(z>0) \in \mathrm{Ref}(z=0)$, blue dashed) are very similar. This indicates that, in these regions, the thermodynamic evolution is largely insensitive to where the particles end up or how hot they become as a consequence of the sub-grid physics.

The cluster plots in Fig.~\ref{fig:medians_tracked_cluster} exhibit trends similar to those in the group, with notable differences. In the core region, the entropy histories of Ref-selected particles in NR ($\mathrm{NR}(z>0) \in \mathrm{Ref}(z=0)$, blue dashed) and Ref ($\mathrm{Ref}(z>0) \in \mathrm{Ref}(z=0)$, red solid) differ at higher redshifts, indicating the presence of feedback activity, consistent with \cite{McCarthy2011}. Temperature and density histories also reflect these trends, with radiative cooling in Ref driving cooling flows that increase gas density (the red solid $\rho(z=0)$ is higher than the blue dashed). In the cluster, unlike the group, the entropy plateau is most manifest in the core; however, the entropy histories in the intermediate region are similar to those in the core. The outskirts show nearly identical median histories for NR-tracked particles selected in NR ($\mathrm{NR}(z>0) \in \mathrm{NR}(z=0)$, blue solid) and Ref ($\mathrm{NR}(z>0) \in \mathrm{Ref}(z=0)$, blue dashed), indicating that thermodynamic evolution in these regions is largely unaffected by sub-grid physics. These observations also align with \cite{McCarthy2011}, highlighting the scale-dependent impact of feedback on the entropy and thermal evolution of the IGM.
Given these considerations, a key result is that the entropy and halo-centric radius of the blue dashed line ($\mathrm{NR}(z>0) \in \mathrm{Ref}(z=0)$) end up above the red line ($\mathrm{Ref}(z>0) \in \mathrm{Ref}(z=0)$) in the region of the entropy core. This behaviour suggests that the core entropy in Ref is high because the hot gas in NR also has high entropy. Therefore, the core-entropy excess in Ref cannot directly be due to AGN feedback, also because the gas sinks to smaller radii instead of being directly heated and buoyantly moving outwards. Secondary feedback-induced shocks could nevertheless have an impact \citep{10.1111/j.1365-2966.2010.17455.x}.

\section{Discussion and conclusions}
\label{sec:discussion-and-conclusions}

We carried out cosmological zoom‐in simulations of a galaxy group of mass $M_{500} = 8.83 \times 10^{12}$~M$_\odot$ and a galaxy cluster of mass $M_{500} = 2.92 \times 10^{14}$~M$_\odot$ at $z=0$ using the SWIFT‐EAGLE reference model. In our mid‐resolution runs \citep[matching the resolution of the EAGLE L100N1504 volume of][]{eagle.schaye.2015} dark‐matter particles have mass $m_\mathrm{DM}=9.82\times10^{6}$~M$_\odot$ and gas particles $m_\mathrm{gas}=1.83\times10^{6}$~M$_\odot$; low-resolution simulations have particle masses 8 times larger; the high‐resolution simulation of the group uses $m_\mathrm{DM}=1.23\times10^{6}$~M$_\odot$ and $m_\mathrm{gas}=2.29\times10^{5}$~M$_\odot$. Our study investigates the entropy evolution of groups and clusters of galaxies, extending our previous research on the entropy excess relative to the self-similar expectation in \cite{Altamura2023} to higher redshifts. In light of recent X-ray observations \citep{2024Galax..12...24E}, the entropy-core problem highlighted in our previous work appears to have emerged due to an observational selection bias (effects on averaged thermodynamic profiles and hot gas fractions are discussed in \citet{2024A&A...687A.238C} and \citet{2024arXiv241116555P}, respectively). By analysing the thermodynamic history of the IGM, we identified three key moments in the evolution of these systems that we link to their entropy structure.

A key result is that entropy plateaus appear to emerge at characteristic halo masses and that their formation is closely linked to AGN feedback: when a system reaches $\sim10^{12}$ M$_{\odot}$, its entropy profile flattens at the virial radius (Fig.~\ref{fig:entropy-profile-scaled}, left column). This transition corresponds to the point where AGN feedback dominates the thermodynamic regulation, as evidenced by the peak in the specific black hole accretion rate (Fig.~\ref{fig:properties-evolution-1}, bottom row). In groups ($\sim10^{13}$ M$_{\odot}$) this plateau extends down to the core-radius, forming a broad region of nearly constant entropy, and at cluster scales ($\sim10^{14}$ M$_{\odot}$) the core itself becomes isentropic (Fig.~\ref{fig:entropy-profile-scaled}, right column).

Then, we analysed the evolution of the entropy distribution by tracking gas particles to reconstruct their median Lagrangian histories and help interpret the origin of the entropy excess relative to the self-similar expectation. The results in Figs.~\ref{fig:medians_tracked_group} and~\ref{fig:medians_tracked_cluster} suggest a picture where (AGN) feedback preferentially removes low-entropy gas from the progenitors before it can fall into the core region, replacing it with hotter and higher-entropy material. This mechanism effectively erases the central entropy gradient, preventing the formation of a well-defined cool core. A striking example of this effect is seen in the core region, where gas tracked in NR simulations retains a power-law entropy distribution, while its counterpart in the Ref run shows a flattened profile at late times. Notably, the halo-centric distance histories (top row of Figs.~\ref{fig:medians_tracked_group} and~\ref{fig:medians_tracked_cluster}) reveal that this redistribution is not simply a matter of local heating. Gas is physically displaced, in agreement with the findings of \cite{McCarthy2011}, altering the overall thermodynamic balance of the system.

We reiterate that the entropy excess seen in our simulated clusters is not unique to EAGLE-like simulation models, but is a common feature of many cosmological simulations. The entropy levels in our clusters at $z = 0$ (Fig.~\ref{fig:entropy-profile-scaled}, bottom row) are consistently higher than observations of like-mass objects by \citet{entropy_profiles_sun2009} and \citep{entropy_profiles_pratt2010}, but align with recent observational results \citep[X-GAP objects show entropy plateaus,][]{2024Galax..12...24E} and trends seen in other cosmological simulations such as IllustrisTNG \citep{2018MNRAS.473.4077P}, SIMBA \citep{2019MNRAS.486.2827D}, and FABLE \citep{2018MNRAS.479.5385H}. Other X-ray observations of galaxy groups highlighted departures from power-laws: ESO 3060170 \citep{2004ApJ...612..805S}; AWM5, which has a clear entropy plateau between 10-50 kpc similar to that of our group (phase C) in the inner region, but recovering a self-similar power-law at larger radii \citep{2009ApJ...694..479B}; Nest 200047, where the entropy profile is a power-law, but 1 dex higher than self-similar expectations with a baryon fraction of 0.15 \citep{2025arXiv250611312M}. Recent results by \cite{2025arXiv250613907E} depicted SDSSTG 4436 as a (fossil) group with a flat inner entropy profile, and evidence of a giant AGN outburst that prevented the formation of a self-regulating feedback cycle - and therefore destroyed the cool core. This scenario is remarkably similar to our group's, whose entropy levels in the IGM are found to be raised by premature feedback at early times (Section~\ref{sec:lagrangian}). On the other hand, the profiles from the group AWM4 \citep{2008ApJ...673L..17G, 2010MNRAS.407..321O} obtained with \textit{Chandra} present a system with high entropy levels, but evidence of little heating.

Focusing again on the observational samples of \citet{entropy_profiles_sun2009} and \citep{entropy_profiles_pratt2010}, our entropy levels exceed most observed values by up to an order of magnitude in group-mass halos and remain systematically elevated by $\approx0.4$ dex in cluster cores, suggesting that AGN feedback, as implemented, is excessively efficient at heating and ejecting gas. This is evident in lower-mass halos, where the gravitational potential is weaker, allowing outflows to more easily expel low-entropy material, as also found in \citet{McCarthy2011} and \citet{lehle_tng_cluster}. The redshift evolution of the baryon fraction (Fig.~\ref{fig:properties-evolution-2}, top row) further supports this: while the cluster maintains a roughly constant baryon fraction, the group exhibits progressive baryon depletion, indicating that AGN-driven outflows are disproportionately affecting smaller systems. The corresponding increase in core entropy reflects a systematic removal of dense, low-entropy gas, leaving behind an IGM dominated by a high-entropy phase.

As commented by \cite{Bahar2024}, the implementation of AGN feedback efficiencies that scale with black hole accretion rate \citep{2013AN....334..394G} may help address this issue by reducing excess heating in groups, though it remains to be determined whether this can be achieved without causing overcooking.

Despite capturing the overall trends in gas fractions, BCG mass and star formation evolution, our simulations do not yet reproduce the observed diversity in entropy core properties, particularly the formation of long-lived cool cores. Observations show that a significant fraction of clusters retain low central entropy ($K_0 \sim 10$ keV cm$^2$), forming well-defined cool cores \citep{entropy_profiles_pratt2010, entropy_profiles_sun2009}. In contrast, our simulated clusters exhibit flattened entropy profiles in and around the core, characteristic of non-cool-core objects (Fig.~\ref{fig:entropy-profile-scaled}, bottom right). However, it is important to recognise that X‑ray–selected samples suffer from a known cool-core bias, preferentially detecting low entropy, centrally concentrated systems, and potentially missing high-entropy groups \citep{2011A&A...526A..79E}. On group scales, especially, this bias means the true distribution of central entropy levels may be skewed; therefore, the entropy-core \textit{problem} may be overstated. Moreover, with only two halos simulated, it is unclear \textit{a priori} whether our particular systems should correspond to observed cool‑core or non‑cool‑core populations. (A larger suite of simulations is required to establish whether the entropy excess we find is a generic feature or reflects the selection of atypical objects.) Moreover, there are no detailed observational results on systems with $M_{500} \approx 9 \times 10^{12}$~M$_\odot$ yet, suggesting that future work should target simulations of objects in the range of few~$\times 10^{13}$~M$_\odot$ to closely match existing observational samples.

Nevertheless, the discrepancy suggests that our AGN feedback implementation might lack the necessary complexity to control and sustain a self-regulated cooling cycle. Observed AGNs operate in episodic bursts, alternating between phases of strong heating and relative quiescence, allowing cooling flows to intermittently re-establish \citep{2017ApJ...843...28M, 2017MNRAS.468.1917R} and potentially form stable cool cores. However, the snapshots of our simulations are not sampled at a high enough frequency to resolve the variability of the cSMBH, usually around or below the kyr timescale; future works should aim to produce this data to test the link between entropy plateaus and BH variability.

Energy injection schemes of AGN feedback are also key to self-regulation dynamics and, therefore, the core entropy levels. Recent work on isolated halos \citep{husko2024_winds} showed that purely thermal isotropic feedback, such as that used in the Ref model, increases the entropy excess in the core and hinders the formation of cool cores, as do purely kinetic jets. However, adopting a hybrid kinetic-thermal scheme appears to preserve cool cores by preferentially removing gas along specific directions while preventing excessive SMBH mass growth. Therefore, refining the AGN modelling and introducing a mixed thermal and jet-like anisotropic mechanisms appear to be the natural next steps, with the knowledge that varying viscosity and artificial conduction implementations in the hydrodynamics scheme can also have a similar overall impact on the IGM. However, it is possible that these changes, alone, cannot help recover power-law-like entropy profiles from an otherwise entropy plateau. Nevertheless, simulations with cosmological accretion produce strikingly different profiles than isolated objects (see Table~\ref{tab:resolutions-comparison-other-works}), so it remains clear that new sub-grid models designed to experiment on the entropy excess should be tested in a cosmological setting. Finally, adjustments to the cooling physics could also help recover power-law-like entropy profiles. For instance, the ROMULUS-C sub-grid model - where metal cooling is switched off above $10^4$ K - appears to preserve cool cores \citep{2019MNRAS.483.3336T} until merging activity destroys them following the predictions of well-known processes. However, \textit{ad hoc} reductions of the cooling rate are to be considered unphysical.

We envisage two potential avenues to increase the diversity in simulated entropy profiles. The first is improving the modelling of multiphase cooling and condensation in the presence of \textit{magnetic fields} \citep[e.g.][]{2025arXiv250213213G}, which can act as an additional channel to inject and transfer energy and influence the gas mixing \citep{2021MNRAS.503.1327E}. Another key improvement would be the incorporation of \textit{mechanical AGN feedback via kinetic jets}, which many observed massive clusters possess. Thermal feedback alone may be too isotropic and efficient at raising entropy, while kinetic outflows, when appropriately calibrated, could preferentially remove gas along certain directions, preserving denser filaments that facilitate cooling.

The details of SMBH seeding could also play an important role. In the lower-resolution BAHAMAS runs \citep{mccarthy_bahamas}, SMBHs were seeded in $10^{11}$ M$_\odot$ halos rather than $10^{10}$ M$_\odot$ as in our simulations, resulting in a later onset of AGN feedback. Therefore, group progenitors in BAHAMAS retain almost their full baryon load by the time they grow to $\sim10^{12}$ M$_\odot$, curbing early low-entropy gas ejection and producing cool cores with realistic gas fractions. By contrast, higher-resolution EAGLE-type simulations like ours do seed SMBHs earlier and in smaller halos, potentially ejecting gas too violently and too early, and hence flattening entropy profiles well above the NR baseline \citep{vkb_2005}. Observationally, measurements of entropy profiles and hot gas fractions for systems of mass $\sim10^{12}$ M$_\odot$ at high redshift would help discriminate SMBH-seeding strategies to adopt in numerical simulations; however, very little is still known about this class of objects.

Upcoming X-ray missions, such as \textit{XRISM} and \textit{Athena}, are expected to produce significantly improved constraints on entropy profiles across redshift and halo masses. A key observational test for our findings will be measuring the radial entropy gradients in cool-core and non-cool-core clusters to determine whether they match the predicted entropy plateaus in simulations (e.g. Fig.~\ref{fig:entropy-profile-scaled}). Additionally, the mass-limited SZ-selected cluster samples, like \textit{Planck} and CHEX-MATE \citep{2021A&A...650A.104C}, with new high-resolution X-ray data (e.g. \textit{eROSITA}) will help mitigate biases inherent to existing X-ray-selected samples, improving the statistical robustness of cool-core fraction estimates \citep{2017ApJ...843...76A, Yuan2020}. The recently introduced X-GAP sample also aims to provide an unbiased catalogue of groups observed by XMM-\textit{Newton} with data out to $r_{500}$; preliminary results \citep{2024Galax..12...24E} have shown a large entropy excess and flat entropy profiles similar to those of our group at $z=0$ (Fig.~\ref{fig:entropy-profile-scaled}), in contrast to the power-law-like ones of X-COP \citep{2022A&A...662A.123E}. This systematic survey has the potential to help gain more clarity on the diversity of entropy profiles, so far emphasised only by individual observations of a few exceptional systems, like the ones discussed above.

Beyond entropy, turbulence, and bulk motion measurements from \textit{XRISM} will help probe the role of IGM mixing. If simulations are over-predicting entropy levels due to excessive AGN heating, these observations should reveal lower turbulent energy fractions in real clusters compared to models, particularly in intermediate-mass systems ($M_{500} \sim 10^{13} - 10^{14}$~M$_{\odot}$). Along these lines, high-resolution simulations should explore alternative entropy transport mechanisms beyond artificial conduction, such as cosmic-ray heating \citep{Ruszkowski2017}, anisotropic conduction \citep{2016ApJ...818..181Y, Pellissier2023}, and SMBH spin-jet coupling \citep{Reynolds2021, husko2024_winds}, which is currently missing in the EAGLE physical models.

Additionally, radio observations (e.g., \textit{LOFAR}, \textit{VLA}, \textit{SKA}) can further help constrain AGN outflow structures and their impact on entropy redistribution, potentially hinting at new ways to model AGNs in simulations. Recently, \cite{2025A&A...694A.232G}, using Magneticum simulations in combination with X-ray and radio observations, showed that the cool-core fractions decrease systematically toward higher-mass clusters, highlighting the need for refined AGN models that dynamically regulate feedback efficiency based on local gas conditions, for simulations and observations to agree.

Recent kinetic Sunyaev-Zeldovich (kSZ) effect measurements \citep{2021PhRvD.103f3513S} indicate that most simulations, except for Illustris \citep{2014MNRAS.444.1518V}, would require significantly stronger AGN feedback to match observed galaxy clustering relations. Results from the FLAMINGO simulations suggest that their fiducial feedback model is insufficiently energetic, and stronger feedback can help match kSZ data and slightly alleviate the $S_8$ tension by reducing small-scale clustering amplitude \citep{2024arXiv241019905M}. However, our results from \citetalias{Altamura2023} and those from FLAMINGO \citep[see Fig. 7 of][]{2024MNRAS.533.2656B} have shown that stronger feedback is undesirable because it increases the core entropy and leads to entropy plateaus, exacerbating the tension with X-ray observations. At present, no feedback prescription in simulations is known to match observational results at all scales -- in the field, $\gtrsim$Mpc, with kSZ and in overdense halos, $\lesssim$Mpc, with X-rays -- simultaneously.

The emergence of entropy plateaus at specific halo masses is a robust, physically motivated feature of structure formation, and has now been observed in low-$z$ groups \citep{2024Galax..12...24E}, who report large entropy excesses and flat cores similar to ours. While many observed clusters still exhibit classic cool-core power-law gradients, recent results of X-GAP and SDSSTG 4436 show that observations too found a diversity of entropy profiles. This underscores the need for AGN feedback models that can reproduce the full spectrum of entropy shapes. In this regard, the manifestation of entropy plateaus in individual systems non-trivially depends on the details of the heating and cooling, which must be carefully controlled and characterised. Therefore, feedback prescriptions must be refined to (i) accurately capture both plateau and power-law behaviours without undermining other key cluster properties, and (ii) remain consistent across sub- and super-Mpc scales, from kSZ to X-ray observables. The next generation of simulations, benefiting from increased resolution and updated sub-grid models, will be key to further understanding these issues, examining the correlations between baryonic physics and cosmology, and helping interpret results from upcoming observational surveys.

\section*{Acknowledgements}
We thank the anonymous reviewer for their comments, which improved the quality of this work.
EA thanks Adrian Jenkins and Alastair Basden for high-performance computing support.
This work used the DiRAC@Durham facility managed by the Institute for Computational Cosmology on behalf of the STFC DiRAC HPC Facility (\href{https://dirac.ac.uk}{www.dirac.ac.uk}). The equipment was funded by BEIS capital funding via STFC capital grants ST/K00042X/1, ST/P002293/1, ST/R002371/1 and ST/S002502/1, Durham University, and STFC operations grant ST/R000832/1. DiRAC is part of the National e-Infrastructure. EA acknowledges the STFC studentship grant ST/T506291/1 and support from the Jodrell Bank Centre for Astrophysics at the University of Manchester. The research in this paper made use of the following software packages and libraries: 
\textsc{Python} \citep{van1995python},
\textsc{Numpy} \citep{harris2020array},
\textsc{Scipy} \citep{virtanen2020scipy},
\textsc{Numba} \citep{lam2015numba},
\textsc{Matplotlib} \citep{hunter2007matplotlib, caswell2020matplotlib},
\textsc{SWIFTsimIO} \citep{Borrow2020},
\textsc{Astropy} \citep{robitaille2013astropy, price2022astropy},
\textsc{Unyt} \citep{goldbaum2018unyt},
\textsc{SWIFT} version 0.9.0 \citep{schaller_2018_swift, 2024MNRAS.530.2378S}, and
\textsc{VELOCIraptor} \citep{2019PASA...36...21E}.

\section*{Data and Code Availability} 
The \textsc{SWIFT} code is licensed under the GNU General Public License v3.0 and publicly available at \href{https://gitlab.cosma.dur.ac.uk/swift/swiftsim}{gitlab.cosma.dur.ac.uk/swift/swiftsim}. The \textsc{VELOCIraptor} structure-finding code is publicly available at \href{https://github.com/ICRAR/VELOCIraptor-STF}{github.com/ICRAR/VELOCIraptor-STF} and distributed under the MIT License. The raw snapshots and halo catalogues can be made available upon reasonable request to the corresponding author. The code used for the analysis and the data products used to generate the figures are documented and made publicly available at the following repository: \href{https://github.com/edoaltamura/entropy-core-evolution}{github.com/edoaltamura/entropy-core-evolution}. 

\bibliographystyle{mnras}
\bibliography{main}

\appendix

\section{Entropy distribution in the NR runs}
\label{app:nr-profiles}
\begin{figure*}
    \centering
    \includegraphics[width=\textwidth]{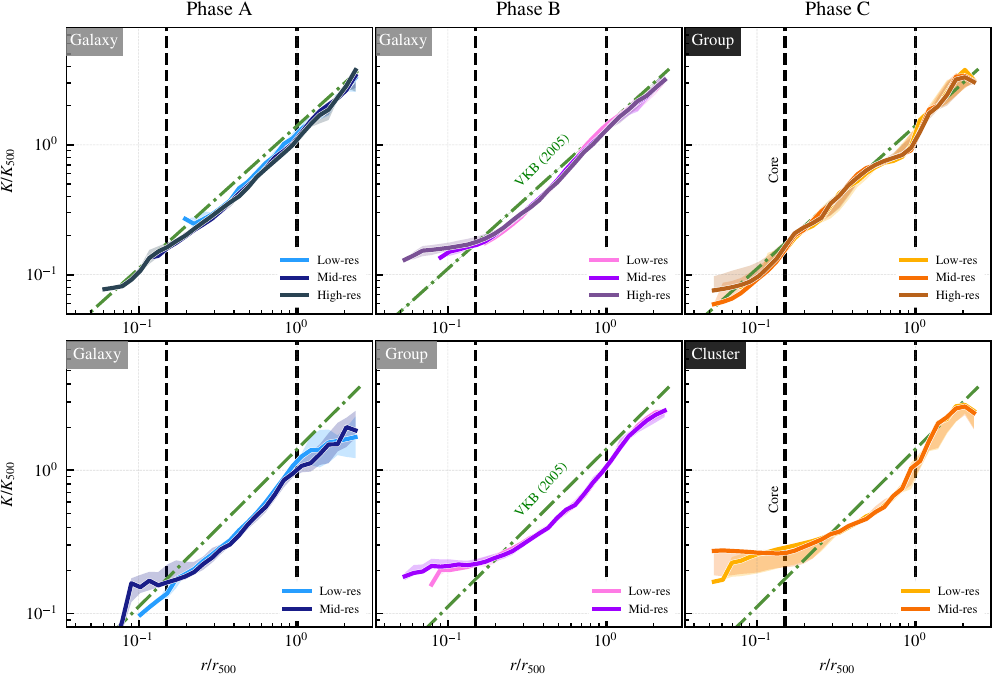}
    \caption{Scaled entropy profiles as in Fig. \ref{fig:entropy-profile-scaled}, but for the NR model.}
    \label{fig:entropy-profile-nr}
\end{figure*}
In addition to the results in Section \ref{sec:results-evolution:entropy-profiles}, we show the scaled entropy profiles for the group and cluster run with the NR configuration, i.e. only gravity and hydrodynamics, in Fig.~\ref{fig:entropy-profile-nr}. The layout is the same as in the Ref profiles: three colours correspond to three evolutionary phases (indicated at the top of the panels), and the hues correspond to runs with different particle-mass resolutions. The NR profiles of both objects are power-law-like down to the core-radius in phases A and B, in line with predictions from \cite{vkb_2005}, shown in green. 

While the NR group maintains a cool core to $z=0$, the cluster's central entropy level departs from the self-similar power law. Given the NR model specifications, the flattening of the cluster's profile can be associated with artificial conduction, which is known to result in the mixing of low- and high-entropy gas. Despite this effect, we note that the entropy level for the cluster profile at the core is $\approx 0.2\, K_{500}$, which is half of that obtained with Ref; moreover, the NR isentropic core extends to $\approx 0.25\, r_{500}$, while the Ref non-cool cores are more extended in size ($\approx 0.5\, r_{500}$).

\bsp	
\label{lastpage}
\end{document}